# Special spectroscopic observations and rare solar events collected in the first half of the twentieth century with Meudon spectroheliograph

J.-M. Malherbe, Observatoire de Paris, PSL Research University, CNRS, LESIA, Meudon, France

Email : Jean-Marie.Malherbe@obspm.fr

ORCID identification : https://orcid.org/0000-0002-4180-3729

## ABSTRACT

Henri Deslandres initiated imaging spectroscopy of the solar atmosphere in 1892 at Paris observatory. He invented, concurrently with George Hale in Kenwood (USA) but quite independently, the spectroheliograph designed for monochromatic imagery of the Sun. Deslandres developed two kinds of spectrographs: the "*spectrohéliographe des formes*", i.e. the narrow bandpass instrument to reveal chromospheric structures (such as filaments, prominences, plages and active regions); and the "*spectrohéliographe des vitesses*", i.e. the "section" spectroheliograph to record line profiles of cross sections of the Sun, in order to measure the Dopplershifts of dynamic features. Deslandres moved to Meudon in 1898 with his instruments and improved the spectral and spatial resolutions, leading to the large quadruple spectroheliograph in 1908 developed with Lucien d'Azambuja. CaII K systematic observations started at this date and were followed in 1909 by Hα with two dedicated 3-metres spectroheliographs. The observing service was organized by d'Azambuja who also intensively used the large 7-metres spectroheliograph for his research and thesis (1930). This paper summarizes fifty years of research by Mr and Mrs d'Azambuja, who explored various photospheric and chromospheric lines, performing special spectroheliograms with the high dispersion 7-metres instrument. They also observed intensively filaments and prominences (memoir published in 1948) and recorded rare solar activity events with the two 3-metres spectroheliographs, during the first half of the twentieth century.

## INTRODUCTION

Henri Deslandres (1853-1948), defended his "*Doctorat ès Sciences*" (1888) in the laboratory of Alfred Cornu at Ecole Polytechnique, and was hired in 1889 by Admiral Ernest Mouchez (the director of Paris observatory since 1878). He was in charge of organizing a spectroscopic laboratory, in the context of the development of physical astronomy initiated by Janssen (1824-1907) at Meudon. Deslandres first built a classical spectrograph to probe the solar atmosphere [note 1] using photographic means; he studied the line profiles of the ionized Calcium at 3934 Å wavelength (the CaII K line) [note 2]; he succeeded to resolve the fine structure of the core in 1892. This was the starting point of the great adventure of spectroheliographs in France. With a classical spectrograph, monochromatic images can be delivered by an output slit located in the spectrum (selecting the light of a spectral line) when the input slit scans the solar surface. This is possible by moving the solar image upon the first slit. The photographic spectroheliograph was invented on this basis by George Hale (1868-1938) in Kenwood (1892) and by Henri Deslandres in Paris (1893), simultaneously but independently. The full story was described in details by Malherbe (2023).

Many spectroheliographs were built in the world, following Hale and Deslandres techniques. For instance, long series of continuous observations in the CaII K line were collected by the spectroheliographs of Kodaikanal in India, Mount Wilson in the USA, Mitaka in Japan, Sacramento Peak in the USA, Coimbra in Portugal, Meudon in France or Arcetri in Italy. They produced extended archives, some of them covering up to 10 eleven-years solar cycles, which are convenient to investigate long-term solar activity and study rare events (such as energetic flares, huge sunspot groups or giant filaments [note 3]). The CaII K3 and Hα central intensities are very convenient for imaging spectroscopy of the chromosphere (filaments, prominences, magnetized bright plages, altitude range 1500-2000 km); the wings of the CaII K line (K1v) and Hα form lower (200-500 km range) and are good tracers of sunspots and faculae [note 3]. The CaII K line is mainly sensitive to temperature fluctuations, while Hα is more sensitive to the density. In parallel, between 1919 and 1939, "section" spectroheliograms of CaII K were systematically performed in Meudon. In that case, instead of moving continuously, the entrance slit of the spectrograph scanned the sun by steps of 20''-25'', forming on the photographic plate full spectra of cross sections of the Sun. The spectral width was about 2 Å, allowing to evaluate the Dopplershifts [note 4] of the CaII K3 component (the line core). Such observations were unique, but the derivation of radial velocities was, of course, a purely manual and long task. For that reason, this program was abandoned in 1939 at the entry of WW2, and never restarted later. However, the 2017 version of Meudon spectroheliograph, with a fast sCMOS numerical detector, registers the full line profiles and provides now 3D FITS data-cubes (x, y, λ), which are available on-line and allow Dopplershifts determination and more parameters using line inversion techniques.

This paper reviews briefly fifty years of monochromatic observations performed in Meudon observatory, during the first half of the twentieth century, under the auspices of Lucien d'Azambuja (1884-1970, figure 1), who retired in 1954. Section 1 describes systematic observations started in 1908. Section 2 relates the studies performed by d'Azambuja and Marguerite Roumens (1898-1985), forming together the famous team of prominence explorers, until 1948. Section 3 summarizes special observations of various spectral lines, done before 1930 for d'Azambuja's thesis with the 7-metres spectroheliograph, which demonstrated that the initial choice of CaII K and Hα was the best for imaging spectroscopy of the chromosphere. At last, section 4 presents rare events of solar activity observed with the more classical 3-metres spectroheliographs.

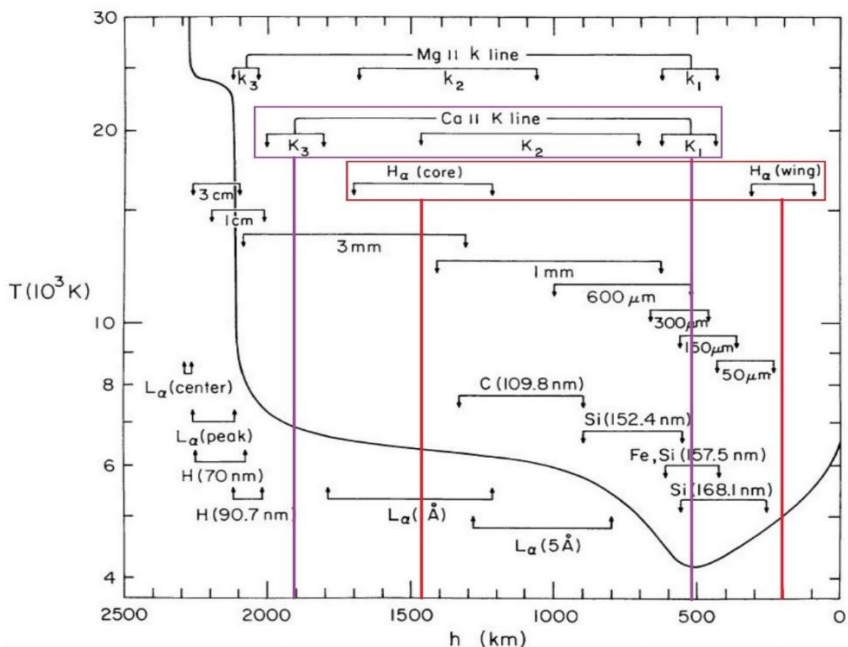

**Figure 1**. *Lucien D'Azambuja (left, courtesy Paris Observatory) is performing observations (right) at the eyepiece of the Hα visual spectrohelioscope, in Meudon (picture of J. Boyer, in "La Nature", n°3198, page 291, 1951).*

## 1 – SYSTEMATIC OBSERVATIONS AT MEUDON

Systematic observations started in 1908 for CaII K and 1909 for Hα with the new spectroheliographs installed in Meudon by Deslandres and d'Azambuja (Deslandres, 1910; d'Azambuja, 1920a, 1920b, 1930).

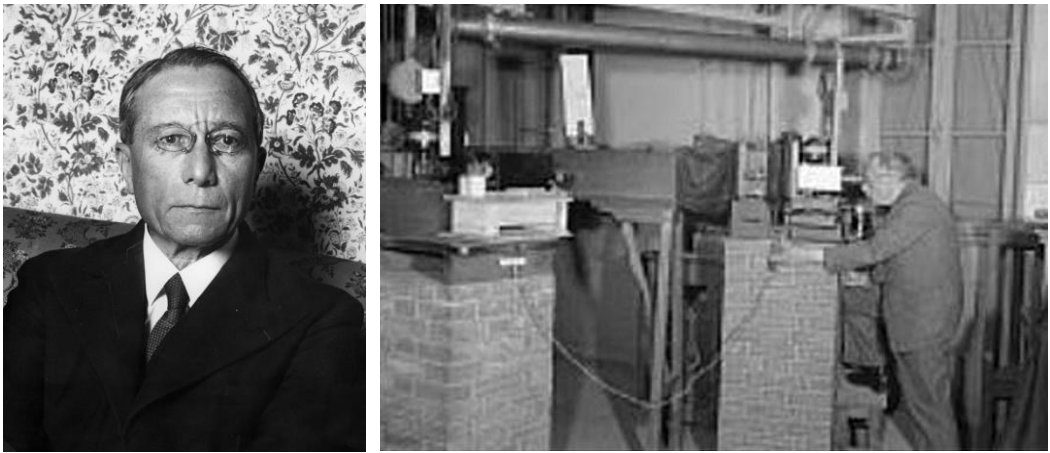

**Figure 2**. *Formation heights (in abscissa) of various solar spectral lines. The cores of Hα and CaII K (named K3) form respectively at about 1500 and 1900 km in the chromosphere (the 500-2100 km layer with increasing temperature). The CaII K2 peaks (K3 ± 0.2 Å) form in the low chromosphere (1000 km), while the wings of Hα and CaII K (named K1, i.e. K3 ± 1.3 Å) form respectively at about 200 and 500 km in the photosphere (the 0-500 km layer with decreasing temperature). The MgII k line in UV (observed by the IRIS spectrograph in space) looks like CaII K. After Vernazza, Avrett & Loeser, 1981, ApJS, 45, 635.*

Systematic observations were performed with two 3-metres spectroheliographs: (1) spectrograph n°I for CaII K using 3 prisms of 60° at minimum deviation, providing a spectral resolution of 0.15 Å, and (2) spectrograph n°II for Hα using a Rowland plane grating, 568 groves/mm, providing in the first order a spectral resolution of about 0.40 Å. The focal length of the scanning objective was 4.0 m, and the magnification of the spectrographs was 2.31 (3.0 m chamber, 1.30 m collimator), so that the size of the Sun on the photographic glass plates was 86 mm. The width of the entrance slit was 0.03-0.04 mm (1.5"-2") well adapted to the usual seeing of 2" in Meudon; while the second slit in the spectrum was 0.075 mm wide.

From 1909 to WW1 (1914), observations consisted of daily spectroheliograms in CaII K3 (line centre, chromosphere), CaII K1v (violet line wing, photosphere) and Hα centre (chromosphere). They were totally interrupted during WW1. Figure 3 shows typical observations performed daily from 1919 to 1939 until the start of WW2. During this period, CaII K3, CaII K1v and Hα were continued, and a "section" spectroheliogram (figure 4) with an enlarged slit in the spectrum (1 mm = 2 Å) was added for Dopplershift measurements of the CaII K line; for that purpose, full spectra of 86 cross sections of the solar surface were recorded by spatial steps of about 22"). Contrarily to WW1, observations were not stopped by WW2, but the lack of manpower was an important constraint, so that "section" spectroheliograms were abandoned, and never restarted after WW2. Typical systematic observations done after WW2 are shown in figure 5; they still include CaII K3, K1v and Hα, plus a long exposure K3 for limb and prominences, with a disk attenuator (because prominences are 5-10 times fainter than the disk).

All data since 1908 are freely available on-line at https://bass2000.obspm.fr, in 12 bits FITS and 8 bits JPEG format for images after 1980, and in JPEG only for older spectroheliograms. The collection is one of the longest world-wide, with more than 100000 images along 10 solar cycles.

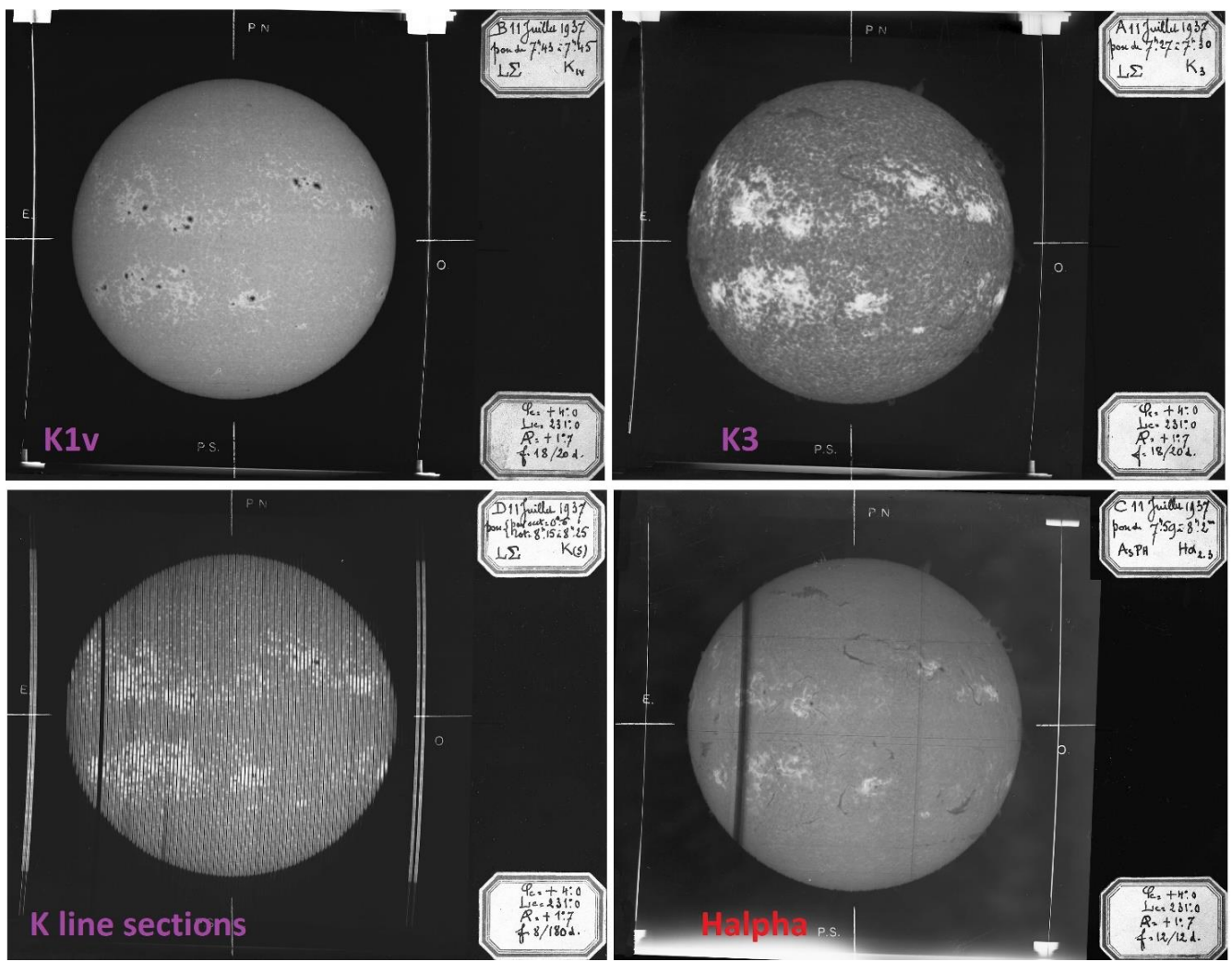

**Figure 3**. Typical *spectroheliograms of the period 1919-1939 (example of 11 July 1937). CaII K1v for the photosphere (sunspots, faculae); CaII K3 for the chromosphere (filaments, bright plages); CaII K line profiles (cross sections of the Sun*

*by 22" steps for Dopplershift measurements); and Hα for the chromosphere (filaments, plages). Courtesy Paris observatory.*

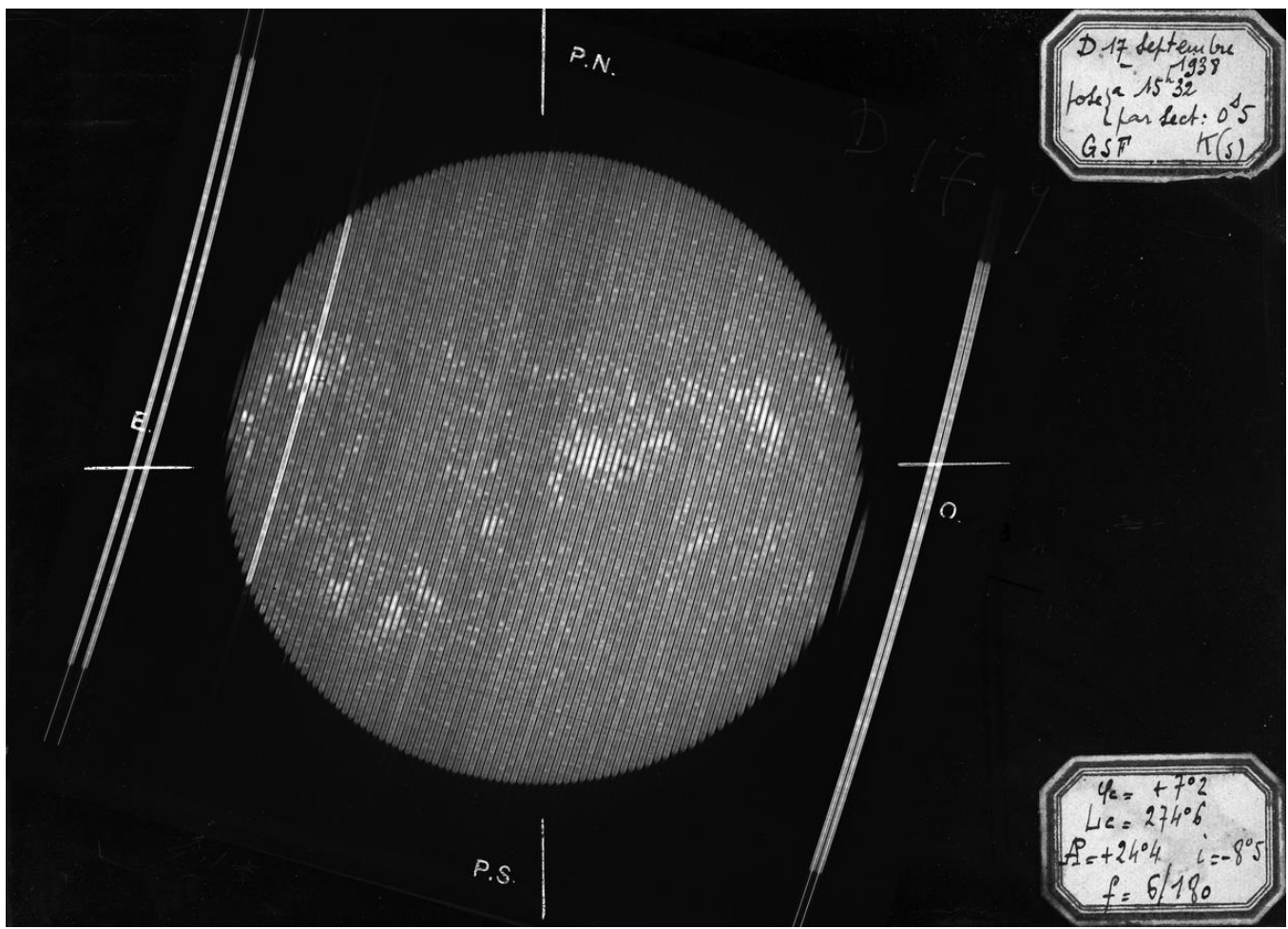

**Figure 4**. *Typical example of "section" spectroheliogram in CaII K line, 17 September 1938. The imaging objective was moving by steps of about 22". The slit in the spectrum was enlarged from 0.075 mm to 1.0 mm, in order to pass 2.0 Å instead of 0.15 Å, the usual bandwidth for CaII K3 and CaII K1v images of the chromosphere. Each of the 86 cross sections provides the line profiles; the Dopplershifts of the line core (K3) were measured and results reported as an index (0-5) in the tables accompanying the synoptic charts of the chromospheric structures, started in 1919 by d'Azambuja for each rotation. As the Dopplershift determination was very time consuming for observers, which were not numerous, this work was done during twenty years (1919-1939) and was abandoned before WW2. Courtesy Paris observatory.*

The spectroheliograph dedicated to daily observations was revisited in the eighties. The prisms (for CaII K) and plane grating at order 1 (for Hα) were replaced by a unique 300 groves/mm grating (blaze angle of 17°27') providing alternatively Hα at order 3 and CaII K at order 5, without any rotation, so that the same 3-metres chamber was used for both lines. The order 4 was convenient for Hβ but the lenses were achromatized only for orders 3 (red) and 5 (violet). Interference orders were selected by coloured filters. Glass plates were abandoned and systematic observations were produced in this new configuration with 13 x 18 cm² film plates until year 2000. Almost real time digitization of films started in 1995 with a 12 bits scanner and images were archived by the BASS2000 data base (https://bass2000.obspm.fr) when it was commissioned in 1996.

The second slit in the spectrum was removed in 2001 when film plates were replaced by a back illuminated CCD camera (1340 x 100 format) from Princeton Instruments (14 bits dynamic range, 20 μm pixels), reducing the chamber focal length from 3.0 m to 0.9 m. Finally, full line profiles are recorded since 2017 with a new scientific CMOS camera from PCO (16 bits dynamic range, 2048 x 2048 format, 6.5 μm pixels) for each location (x, y) of the solar surface, with a chamber reduced to 0.40 m only. As CaII H and CaII K are simultaneously observed with this sCMOS array, both lines are recorded (see movie 1 and movie 2 for CaII K and Hα data-cubes). The spectral sampling of CaII H and K is 0.093 Å (order 5), while the one of Hα is 0.155 Å (order 3). The spectral domain of usual observations is 9 Å and 6 Å, respectively for CaII H, K, and Hα.

The spatial sampling is about 1.1", which is quite convenient for the standard Meudon seeing of 2". The width of the entrance slit of the spectrograph never changed (0.03 mm) and corresponds to 1.55" on the solar surface.

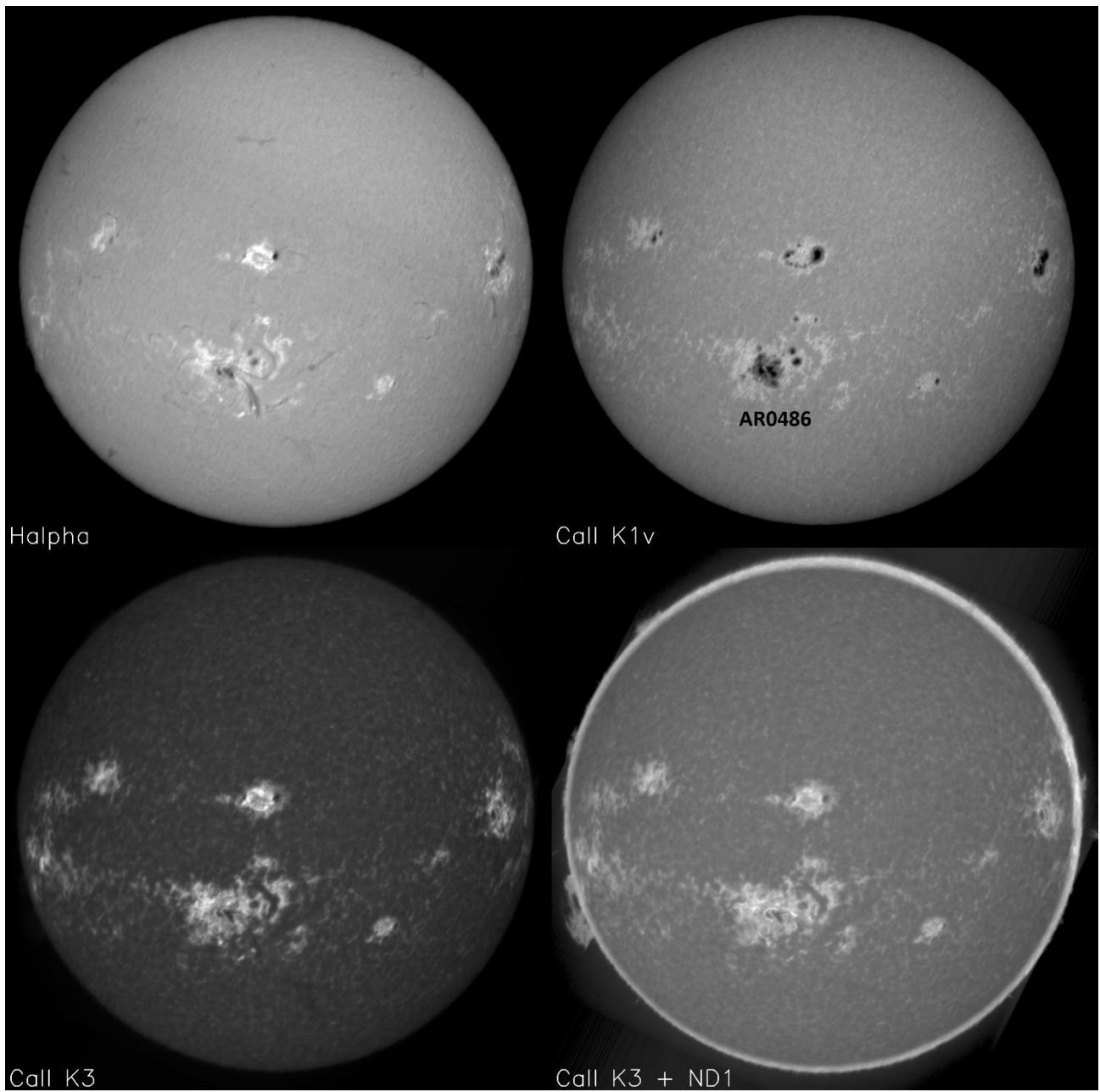

**Figure 5**. *Typical spectroheliograms after WW2 (example of 28 October 2003, the day of a massive X-class flare). Hα for the chromosphere (filaments, plages); CaII K1v for the photosphere (sunspots, faculae); CaII K3 for the chromosphere (filaments, bright plages); CaII K3 long exposure time for prominences and limb with a neutral density attenuator ND1 (10% transmission) superimposed on the solar disk. Courtesy Paris observatory.*

## 2 - Mr AND Mrs d'AZAMBUJA, THE TWO FAMOUS PROMINENCE EXPLORERS !

D'Azambuja was hired in 1899 by Deslandres when he was only 15 years old (Martres, 1998). Of course, he completed in parallel his studies in mathematics and physics at Paris University. Marguerite Roumens arrived at Meudon in 1925 and participated to the spectral line observing program of d'Azambuja (section 3) with the 7-metres spectroheliograph; the original results were reported in his thesis (d'Azambuja, 1930). They got married and started a common and monumental research work about solar filaments (d'Azambuja & d'Azambuja, 1948). This memoir is considered as a "bible" among Meudon solar astronomers. It contains an extensive study of the morphology, evolution and stability of solar filaments and prominences ([figure 6](#)), and their relationships with their neighbourhood (active regions). It is a funding work at the base of modern solar physics. When d'Azambuja retired in 1954 (he was 70 years old), his wife became responsible of observations; she retired in 1959 and the d'Azambujas left together Meudon at this date, after 60 years of astronomy for Mr d'Azambuja ([Malherbe, 2023](#)) ! An incredible career, impossible to reproduce today…

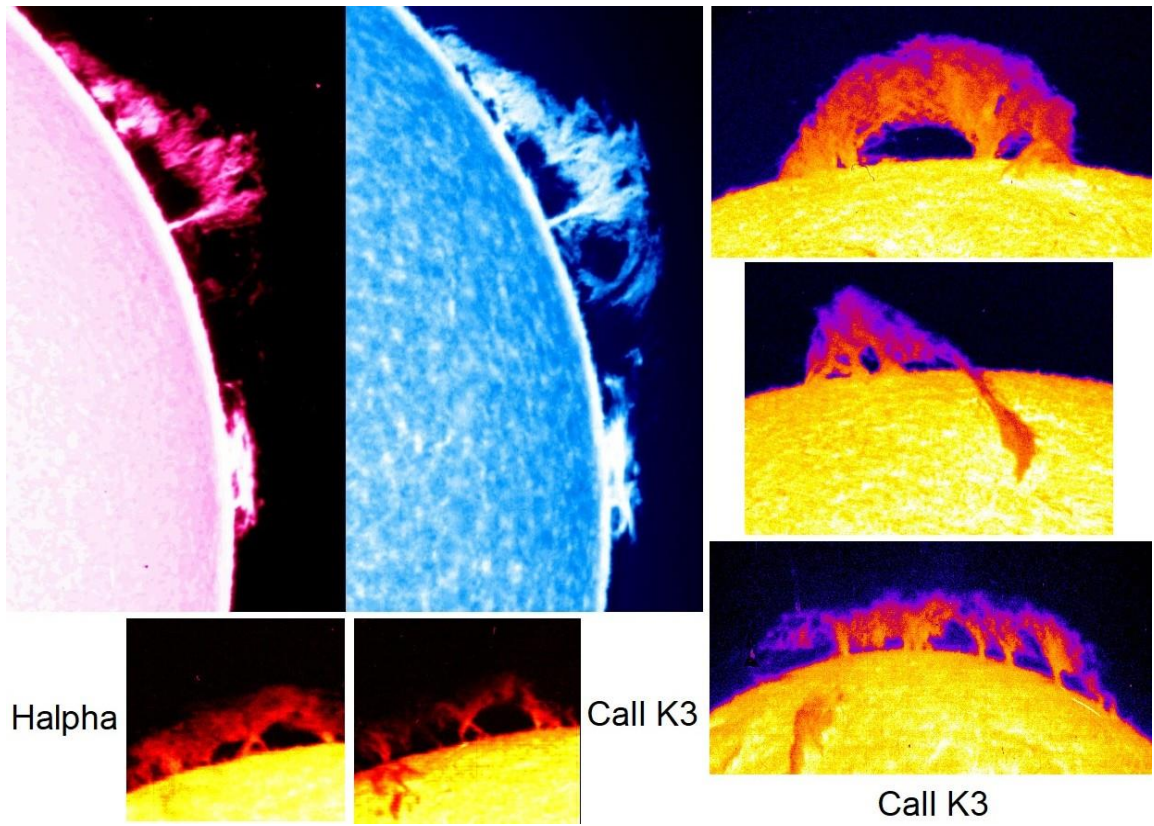

**Figure 6**. *Various aspects of filaments and prominences in Hα and CaII K, observed with Meudon spectroheliographs during the first half of the twentieth century (false colours). Courtesy Paris observatory.*

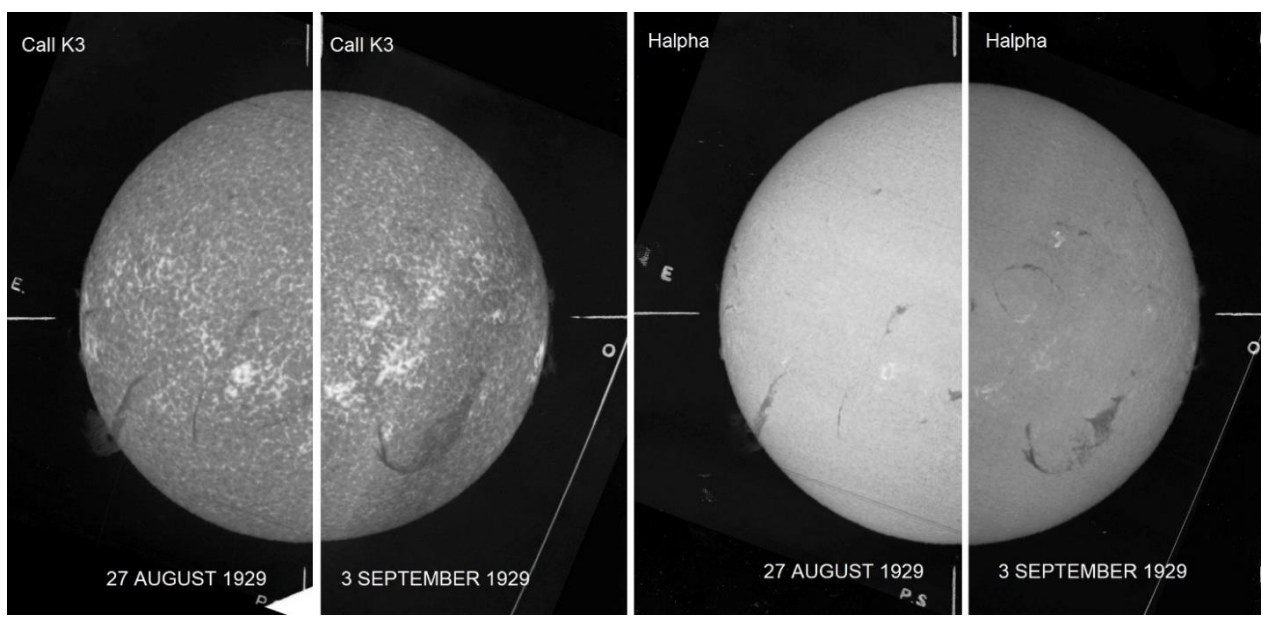

**Figure 7**. *A solar filament exhibiting various perspective effects during the rotation (27 August/3 September 1929), in CaII K3 (left) and Hα (right). Courtesy Paris Observatory.*

This huge study of 387 filaments showed that they are thin (5000 km typical), long ($10^5$ km to $10^6$ km), high (up to 50000 km) and cold (8000 K) Hydrogen structures suspended in the hot corona ($10^6$ K), above the chromosphere. Filaments look like a vertical blade anchored into the photosphere by legs and barbs on both sides. They exhibit extremely variable aspects due to perspective effects during the solar rotation (figure 7 and figure 8). When seen at the limb, filaments are called bright prominences which emit light in Hα or CaII K. But when seen on the solar disk, filaments absorb the photospheric light in the same lines and appear dark.

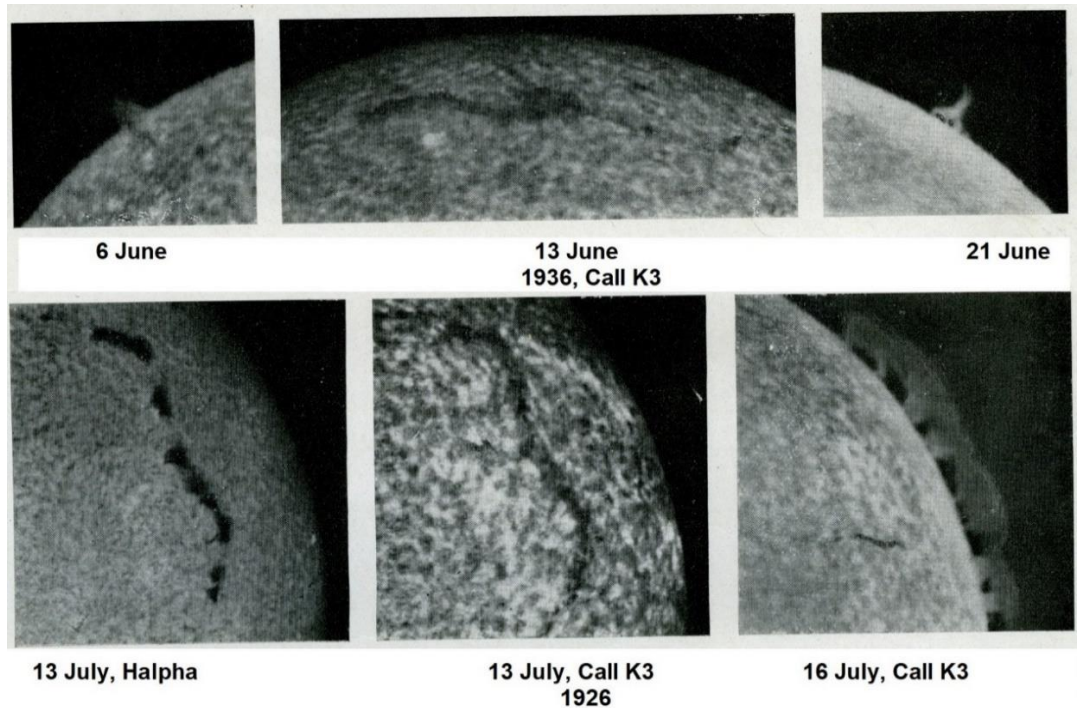

**Figure 8**. *Various aspects of solar filaments during the rotation. Emitting light at the limb (bright prominences), or absorbing light on the disk (dark filaments). After d'Azambuja & d'Azambuja (1948).*

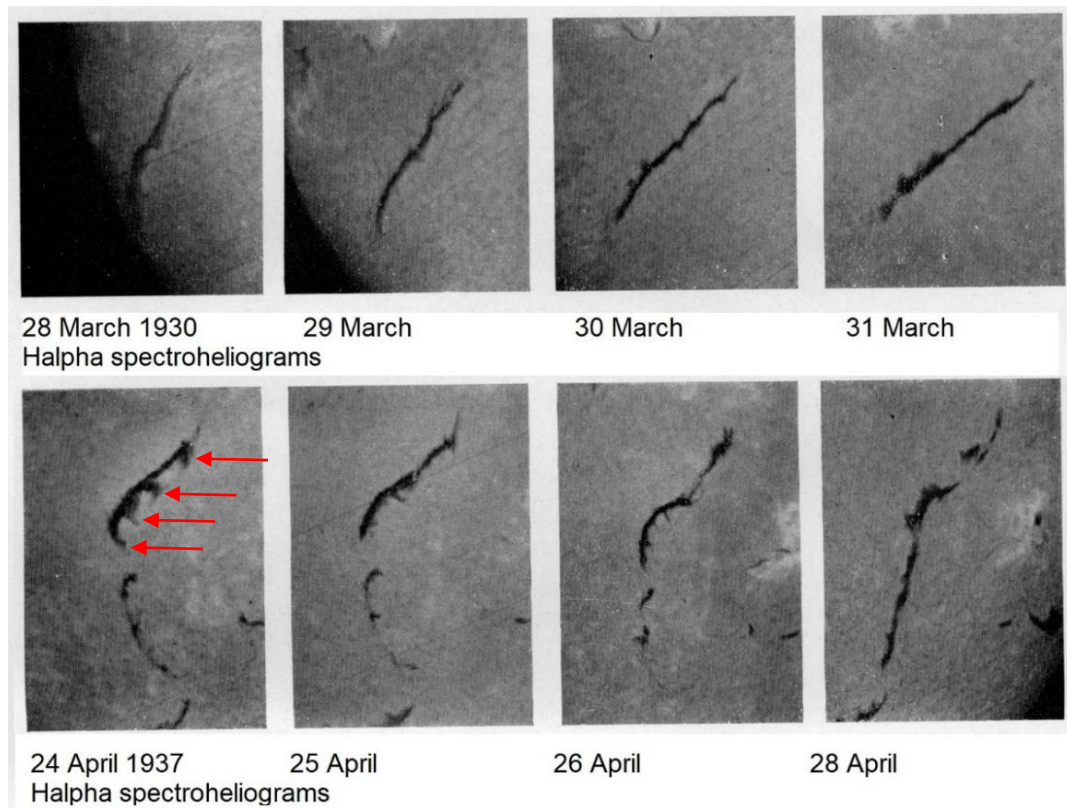

**Figure 9**. *Evolution of filaments and their anchorage into the photosphere (arrows indicate barbs along the filament and legs at both ends). After d'Azambuja & d'Azambuja (1948).*

Figure 9 shows a typical example of filament evolution during a few days. D'Azambuja & d'Azambuja (1948) found that quiescent filaments (far from active regions) evolve slowly and can last during several rotations (the duration of the synodic or Carrington rotation is 27.27 days). Their shape and length vary continuously, filaments can split into several parts, which can also reconnect. Varying perspective effects play a major role during the solar rotation. The authors found a systematic inclination (8°) of the filament blade towards the western direction. The orientation of equatorial filaments is not far from the meridians (figure 9), while polar ones, which appear at the beginning of a new cycle, are close to the parallels (figure 10). The differential rotation (faster at low latitudes) affects the direction of long filaments and elongate them. We know today that filaments are dense and cold structures (8000 K) supported in the tenuous and hot corona ($10^6$ K) by weak magnetic fields (20 Gauss) overlying the boundary between regions of opposite polarities. At the beginning of a new cycle, polar filaments delineate the frontier between the new emerging magnetic field at lower latitudes and the old polar one which reverses at the next solar maximum. The authors emphasized the importance of royal zones, both for the birth of active regions and filaments; they discovered that filaments migrate towards the poles, indicating the existence of a meridional circulation, which was confirmed later by helioseismology (the MDI instrument onboard SOHO/ESA/NASA).

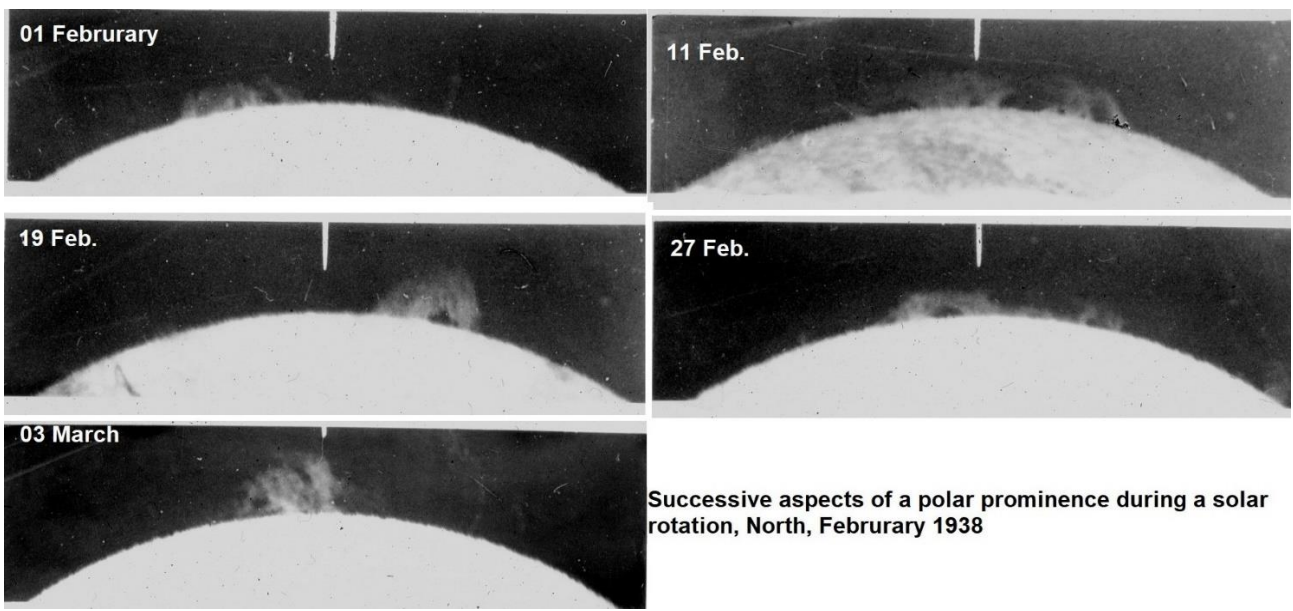

**Figure 10**. *Evolution during the solar rotation of a polar prominence (February 1938). Father Angelo Secchi (1818-1878) wrote in his book "Le Soleil": "Prominences exhibit so various shapes that it is impossible to describe them with any exactitude"… Courtesy Paris Observatory.*

The rotation of filaments was found similar to the one of active regions and sunspots, except at high latitudes where measurements revealed a slower rotation. Among the 387 studied filaments, 141 were associated to active regions (plage filaments), while 246 were not (quiescent ones). Their temporal evolution was found similar: they need (in average) 3 rotations to develop, and start to fragment during the fourth passage on the disk. They can last up to 6 rotations (even 20 rotations for very quiet polar prominences !) and their length increases with their lifetime. Some filaments were observed to split or merge together. They may become unstable. During a full solar cycle (11 years), D'Azambuja & d'Azambuja (1948) found that 206 objects partly disappeared (this was called the "Disparition Brusque" phenomenon, or DB) ; while in 137 cases, the DB was only temporary. Most perturbations occured among equatorial filaments close to active regions. We know today that the DB events can be either thermal (heating of material which vanishes in Hα) or dynamic (magnetic instability affecting the filament support, sometimes associated to flares or coronal mass ejections).

## 3 - SPECIAL SPECTROHELIOGRAMS AND UNUSUAL SPECTRAL LINES

During his thesis work, that he defended in 1930, d'Azambuja explored many spectral lines with the large 7-metres spectroheliograph for research, providing a much higher dispersion that the two 3-metres spectroheliographs used for systematic observations (figure 11). High dispersion is required to study thin photospheric lines such as MgI 3838 Å, SrII 4078 Å, FeI 4202 Å, CaI 4227 Å, FeI 4384 Å, MgI 5184 Å, HeI 5876 Å, NaI 5890 Å, as well as chromospheric infrared lines of CaII 8498 Å, CaII 8542 Å, HeI 10830 Å. The large spectroheliograph (dismantled in the sixties) was able to offer the dispersion of 1 mm/ Å, allowing spectral resolutions on line profiles as low as 0.05 Å. Prisms were used in the blue part of the spectrum, and a plane grating for longer wavelengths.

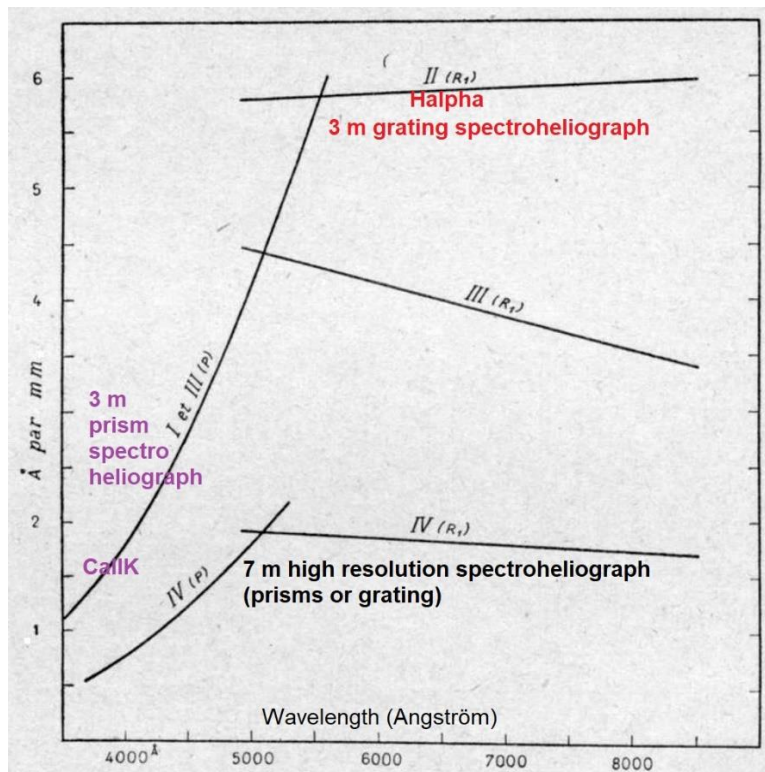

**Figure 11**. *Dispersion of the various spectroheliographs. 3-metres spectrographs I (prisms) and II (grating) were used only for systematic observations, respectively of CaII K and Hα lines. The high dispersion 7-metres spectroheliograph IV (3 prisms or grating) was used only for research work. After D'Azambuja (1930).*

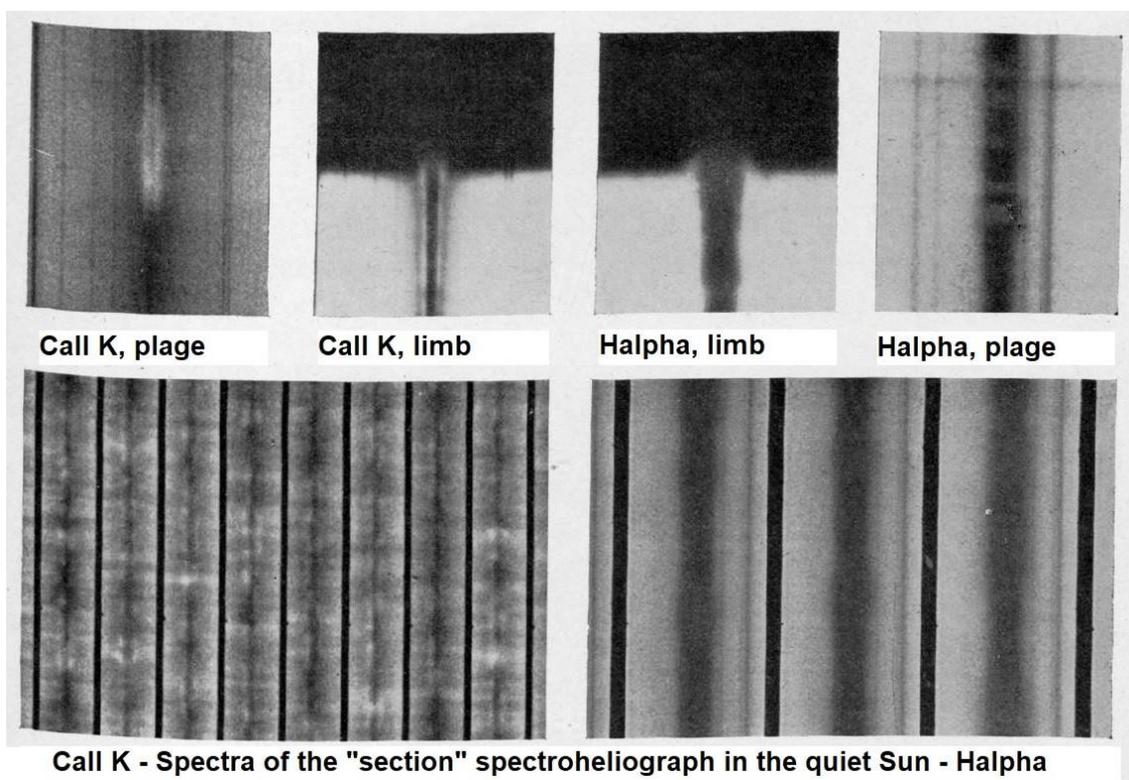

**Figure 12**. *Spectral lines of CaII K and Hα above a bright plage (top), at the limb (top) or in the quiet Sun (bottom). Wavelength in abscissa. After d'Azambuja (1930).*

For instance, the high dispersion 7-metres spectroheliograph, on CaII K and Hα lines, provided the results of figure 12, with a lot of details along the profiles. In plages or at the limb, the bright K2v and K2r peaks (± 0.2 Å apart from the dark K3 core) are well resolved. The chromosphere appears clearly at the limb in both lines above the photosphere (visible in the line wings and the continuum).

Figure 13 displays three images, respectively in the centre of Hα line and in the blue wing. Filaments are still visible at -0.25 Å, but vanish at -0.50 Å (near the inflexion points of the line) and the photosphere below the chromosphere begins to appear at -0.50 Å, except in the case of Dopplershifts (0.5 Å = 25 km/s). Such observations are interesting because they were never done daily, contrarily to the Hα centre.

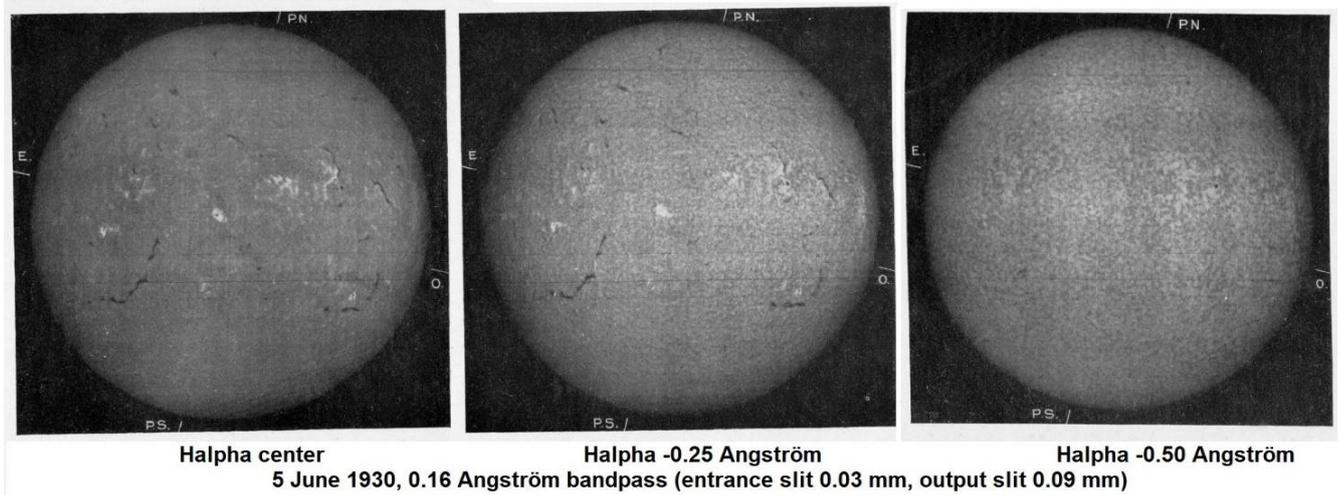

**Halpha center**     **Halpha -0.25 Angström**     **Halpha -0.50 Angström**
**5 June 1930, 0.16 Angström bandpass (entrance slit 0.03 mm, output slit 0.09 mm)**

**Figure 13**. *Observations at three positions in the Hα line. Filaments vanish in the line wing, except in the case of Dopplershifts, while photospheric structures begin to appear. 5 June 1930. After d'Azambuja (1930).*

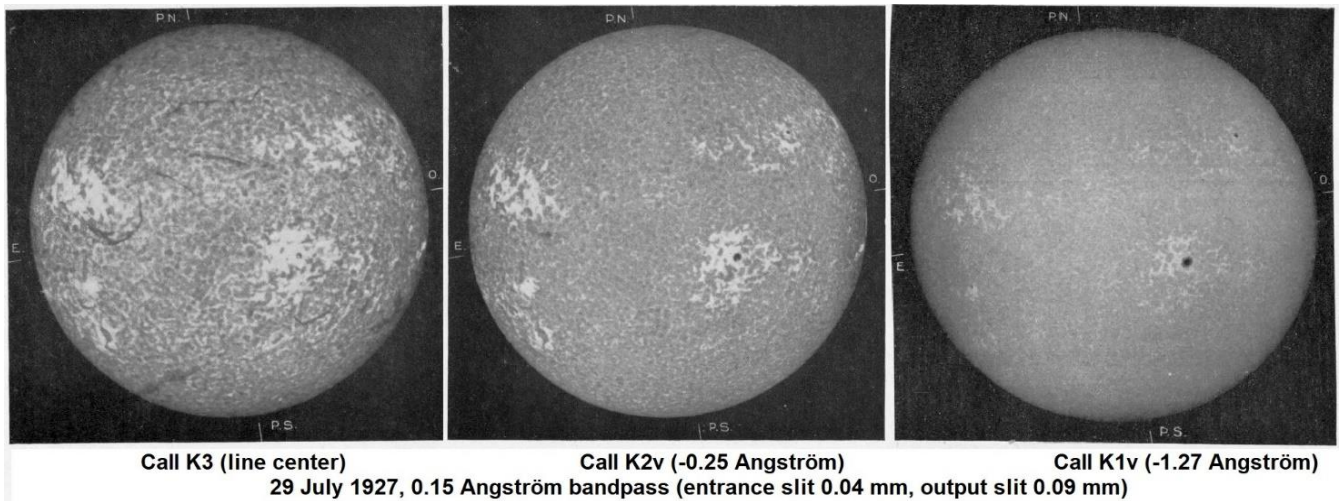

**CaII K3 (line center)**     **CaII K2v (-0.25 Angström)**     **CaII K1v (-1.27 Angström)**
**29 July 1927, 0.15 Angström bandpass (entrance slit 0.04 mm, output slit 0.09 mm)**

**Figure 14**. *Observations at three positions in the CaII K line (K3, K2v, K1v). 29 July 1927. After d'Azambuja (1930).*

Figure 14 displays three images, respectively in the centre of CaII K line (K3) and in the blue wing. Filaments and bright plages are perfectly visible in K3, but this is not the case of sunspots. Filaments are fairly observable at -0.25 Å (K2v) and sunspots begin to appear. But filaments completely vanish at -1.27 Å (K1v) while the photosphere, below the chromosphere, becomes clearly visible, with sunspots and bright faculae (located below chromospheric plages). Observations of K2v are particularly rare in the collection, contrarily to K3 and K1v. Figure 15 shows an exceptional observation of the Balmer series, Hα, Hβ, Hγ, Hδ and Hε. The contrast of filaments and plages (chromosphere) vanishes with decreasing wavelength, because lines form deeper. On the contrary, sunspots appear progressively along the series, which reveals that Hα is the best choice for chromospheric structures, while Hδ or Hε are not better than CaII K1v for sunspots and faculae (however Hε is contaminated by the red wing H1r of the CaII H line). This observational series is the only one of the collection. Figure 16 focuses on the central part of figure 15 with active regions and filaments.

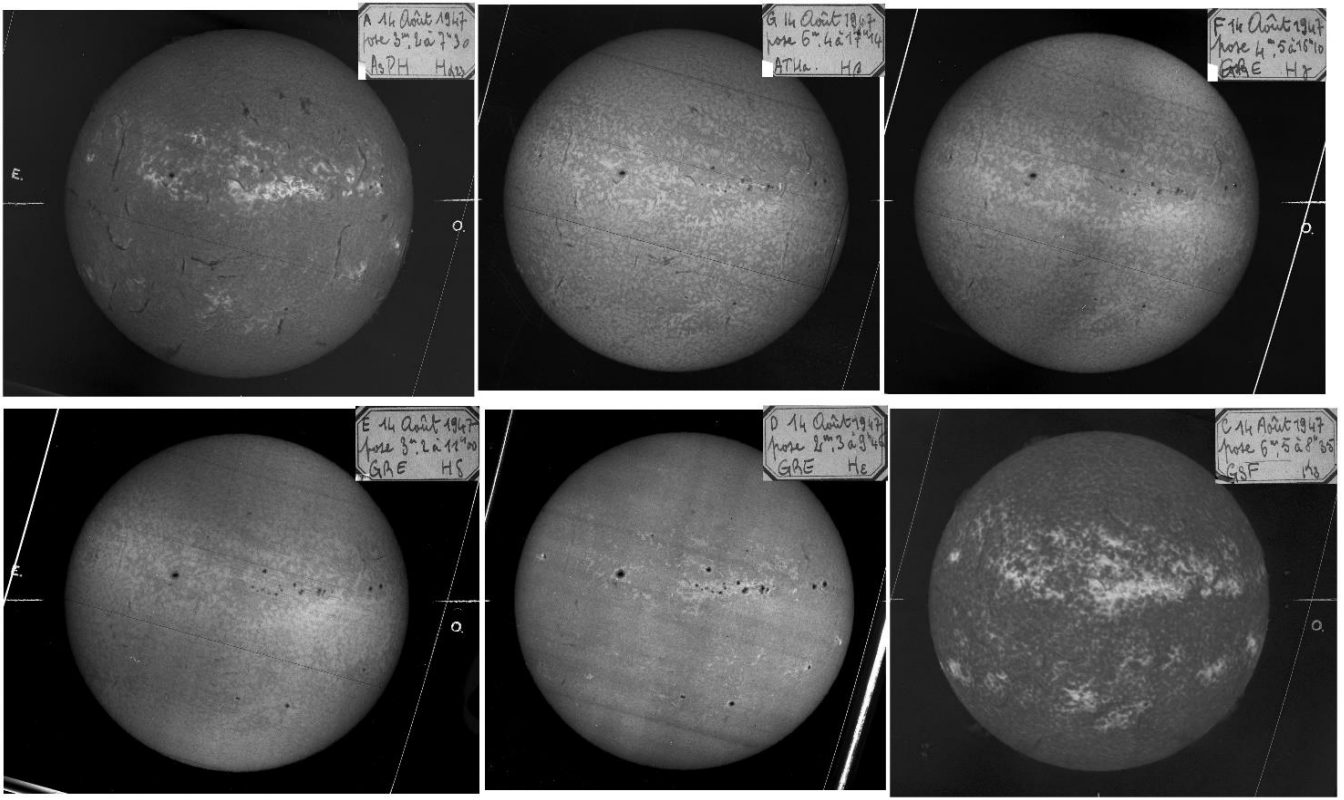

**Figure 15**. *The Balmer series Hα, Hβ, Hγ, Hδ and Hε, 14 August 1947, and the CaII K3 spectroheliogram for comparison. Courtesy Paris Observatory.*

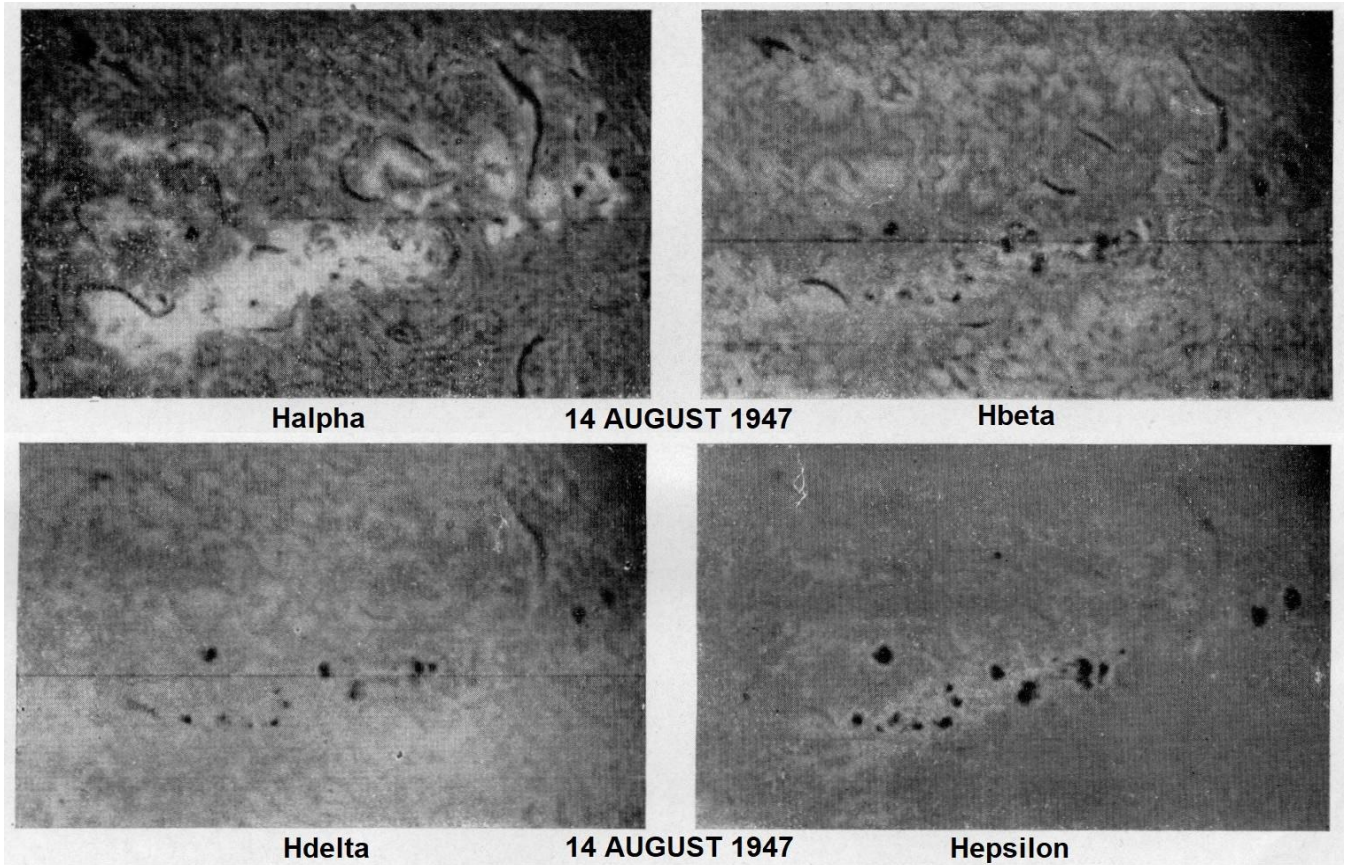

**Figure 16**. *The Balmer series Hα, Hβ, Hδ and Hε, 14 August 1947 (detail of figure 15). The contrast of filaments and plages is the best in Hα and decreases drastically along the series. After d'Azambuja & d'Azambuja (1948).*

Helium lines where also tested by d'Azambuja. Figure 17 shows an exceptional observation in HeI D3 5876 Å of an active region producing the huge flare of 26 July 1946, while figure 18 shows the first world-wide observation done in the infrared HeI 10830 Å line. These observations are absolutely unique in the collection (d'Azambuja, 1938). Several decades later, Kitt Peak National Observatory started systematic observations in the infrared line.

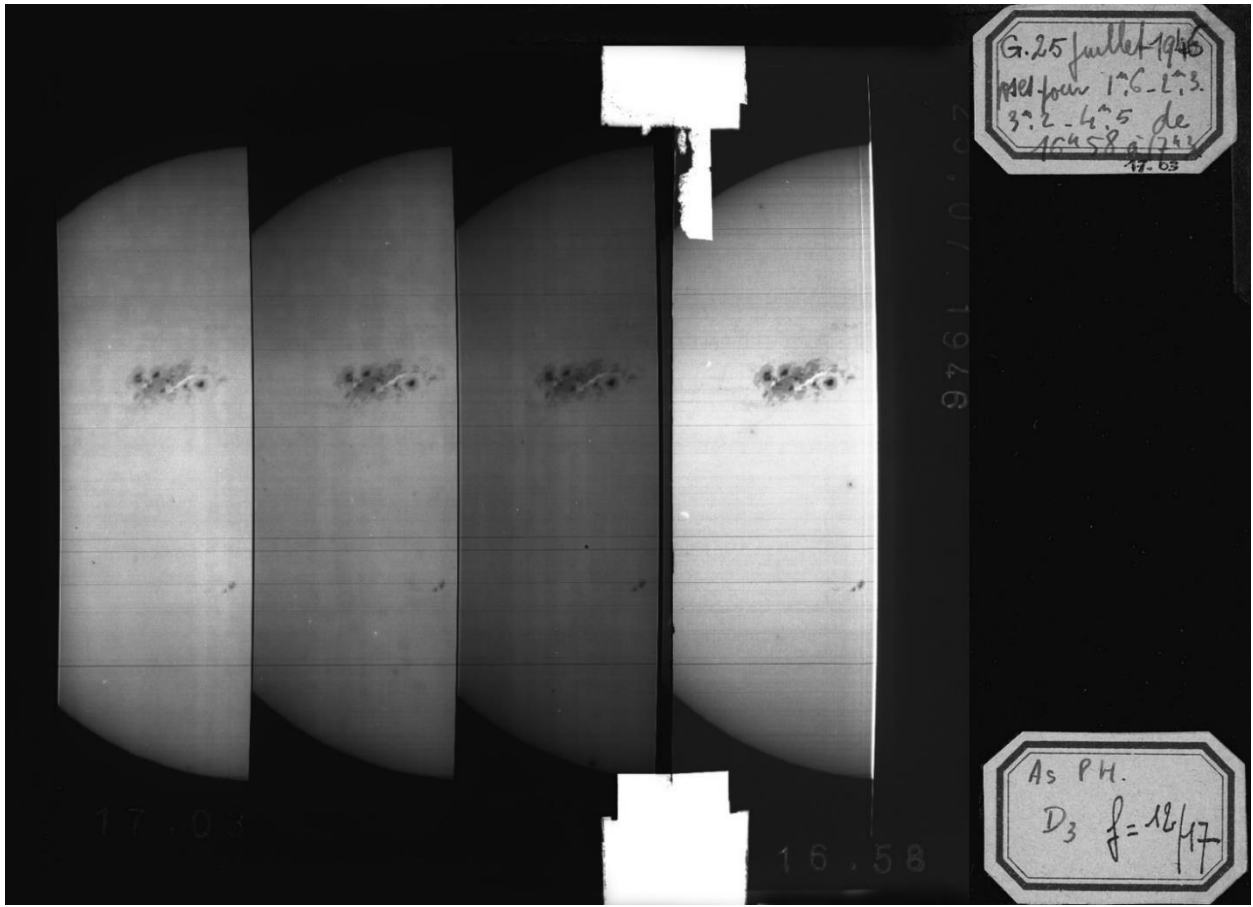

**Figure 17**. *The flare of 25 July 1946 observed in HeI D3 at 5876 Å from 16:58 to 17:03. Courtesy Paris observatory.*

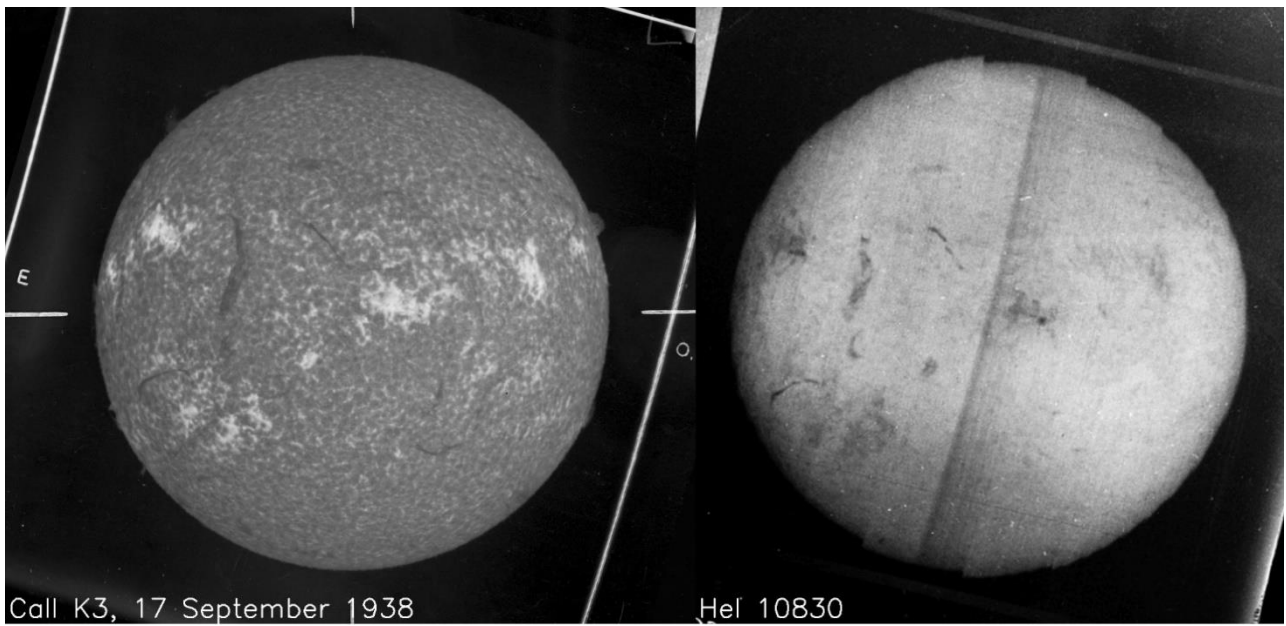

**Figure 18**. *First observation of the infrared HeI 10830 Å line (right), where both filaments and plages appear darker than the quiet Sun. CaII K3 (left) is given for comparison. 17 September 1938. Courtesy Paris observatory.*

Infrared lines of CaII where studied by d'Azambuja in his thesis work. Figure 19 shows images in CaII 8498 Å and 8542 Å, on 4 September 1928, in comparison with the usual CaII K3 3934 Å. These observations are extremely rare in the collection and showed that chromospheric structures are better seen with CaII K3 (figure 20). Several decades later, Kitt Peak National Observatory started systematic observations in CaII 8542 Å in order to measure chromospheric magnetic fields (which is much more difficult with CaII K).

Figure 20 shows also a detail of figure 18, for comparison of HeI 10830 Å and CaII K. While filaments appear dark in both lines, this is not the case of plages, which are bright in CaII K but dark in HeI 10830 Å.

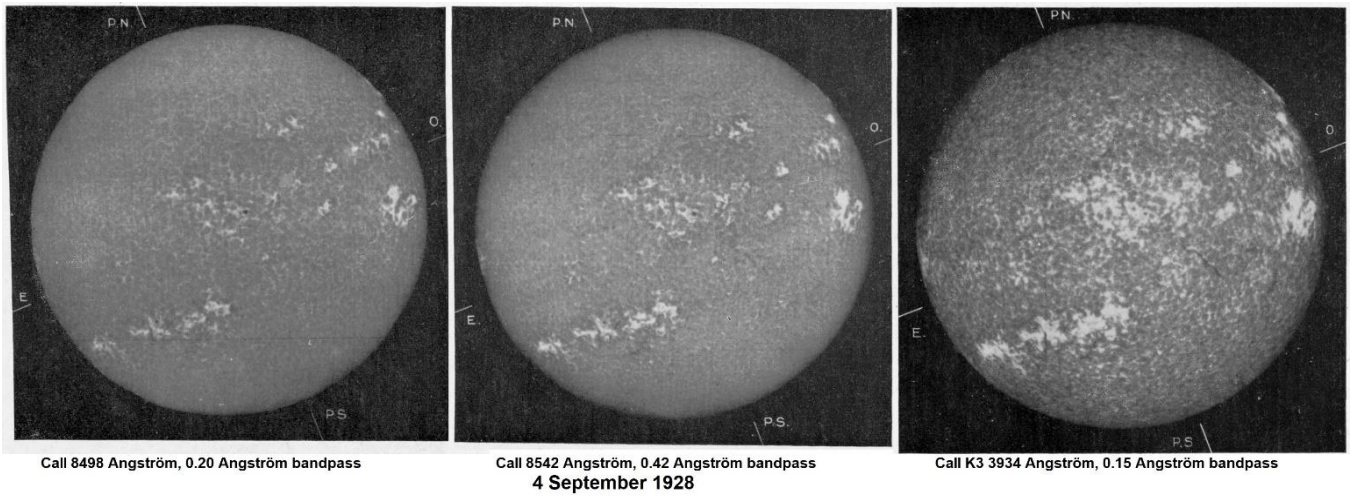

CaII 8498 Angström, 0.20 Angström bandpass    CaII 8542 Angström, 0.42 Angström bandpass    CaII K3 3934 Angström, 0.15 Angström bandpass
**4 September 1928**

**Figure 19**. *CaII 8498 Å and CaII 8542 Å, compared to CaII K3 3934 Å (right). 4 September 1928. After d'Azambuja (1930).*

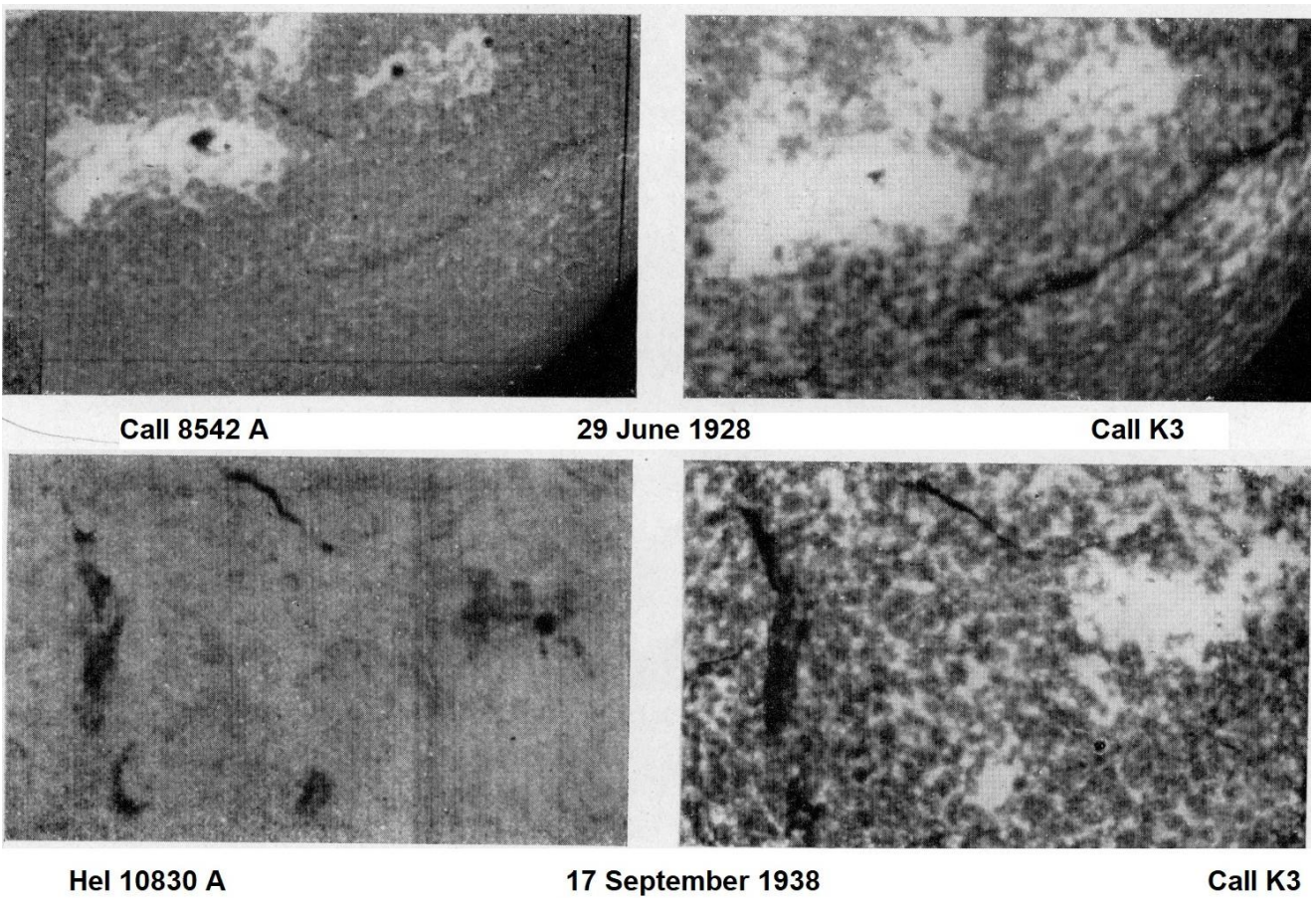

**CaII 8542 A**          **29 June 1928**          **CaII K3**

**HeI 10830 A**          **17 September 1938**          **CaII K3**

**Figure 20**. *CaII 8542 Å, compared to CaII K3 3934 Å (top). 29 June 1928. After d'Azambuja (1930). HeI 10830 Å, compared to CaII K3 3934 Å (bottom). 17 September 1938. Courtsey Paris Observatory.*

CaI 4227 Å line was tested by d'Azambuja during the course of his thesis. Figure 21 shows images in the line core and wing, in comparison with the usual CaII K1v for the photosphere. Figure 22 shows spectroheliograms in FeI 4046 Å, FeI 4202 Å, SrII 4078 Å lines. Many more photospheric lines (list of figure 22 indicating also the selected bandwidth in Å) were observed at high spectral resolution and a facular visibility index (0-5) was established. It sounded clearly that the investigated lines provided spectroheliograms that are comparable to CaII K1v, so that observations in these wavelengths were never reproduced later.

**17 May 1927**

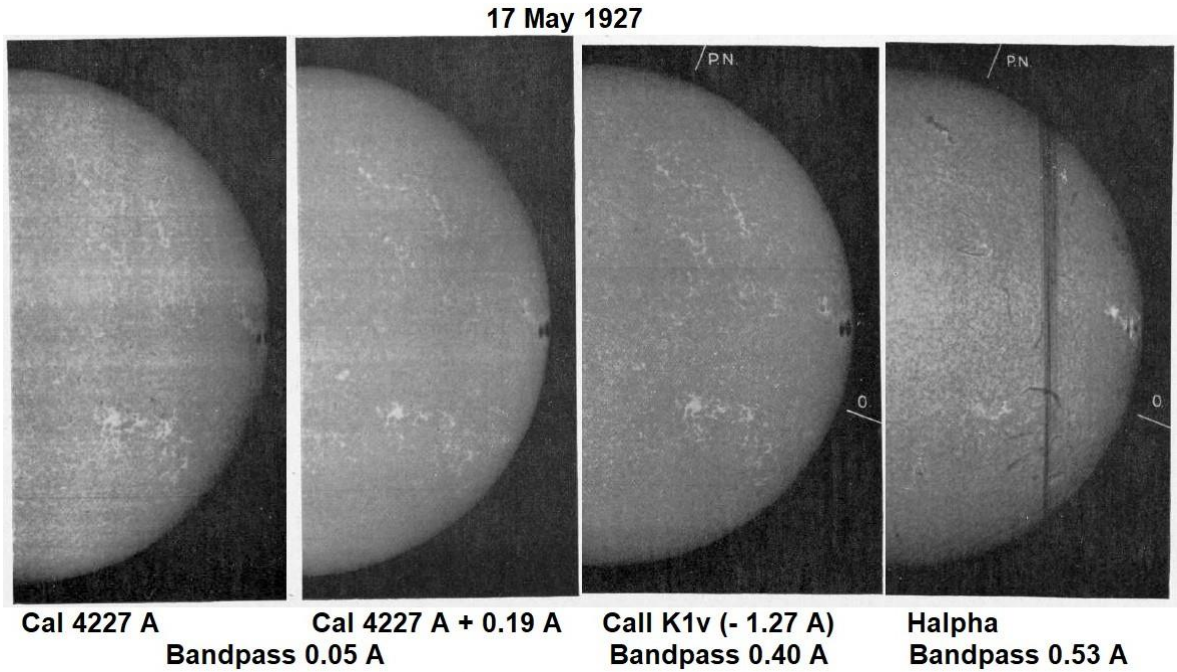

| CaI 4227 A | CaI 4227 A + 0.19 A | CaII K1v (- 1.27 A) | Halpha |
|---|---|---|---|
| Bandpass 0.05 A | Bandpass 0.40 A | Bandpass 0.53 A | |

**Figure 21**. *Spectroheliograms of the CaI 4227 Å line, compared to CaII K1v for the photosphere (at right, for reference, the chromospheric Hα line). After d'Azambuja (1930).*

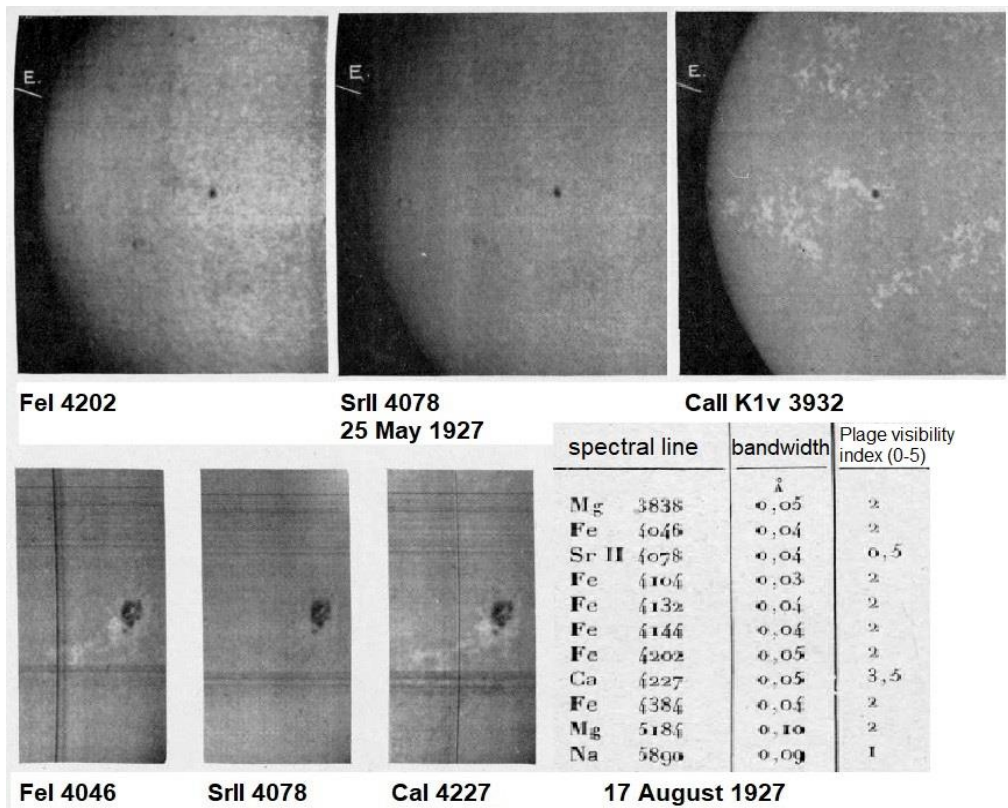

| spectral line | | bandwidth Å | Plage visibility index (0-5) |
|---|---|---|---|
| Mg | 3838 | 0,05 | 2 |
| Fe | 4046 | 0,04 | 2 |
| Sr II | 4078 | 0,04 | 0,5 |
| Fe | 4104 | 0,03 | 2 |
| Fe | 4132 | 0,04 | 2 |
| Fe | 4144 | 0,04 | 2 |
| Fe | 4202 | 0,05 | 2 |
| Ca | 4227 | 0,05 | 3,5 |
| Fe | 4384 | 0,04 | 2 |
| Mg | 5184 | 0,10 | 2 |
| Na | 5890 | 0,09 | 1 |

FeI 4202 — SrII 4078 — CaII K1v 3932
25 May 1927

FeI 4046 — SrII 4078 — CaI 4227 — 17 August 1927

**Figure 22**. *Spectroheliograms in various photospheric lines, compared to CaII K1v. After d'Azambuja (1930).*

# 4 - RARE EVENTS OF SOLAR ACTIVITY OBSERVED WITH THE SPECTROHELIOGRAPHS

The centennial Meudon collection of CaII K1v spectroheliograms contains plenty of information about the area covered by sunspots. In particular, the April 1947 sunspot group (figure 23, centre) is probably the largest ever seen (6.1% of the solar disk), followed by other groups of the same solar cycle (number 18), among them the February 1946, May 1951, July 1946 (figure 23) and March 1947 groups (respectively 5.2%, 4.9%, 4.7% and 4.6% of the disk). These observations illustrate the interest for long term observations spanning more than 10 solar cycles, because exceptional area sunspots provide an upper limit for magnetic energy that can be released in solar flares. Please refer to Malherbe (2022) for more details on such events.

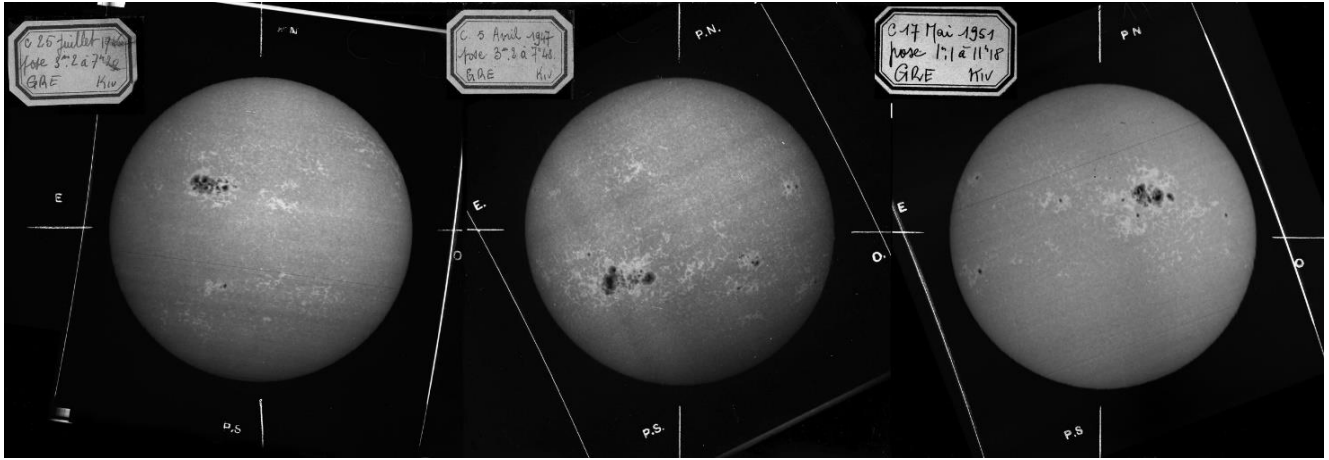

**Figure 23**. *Huge sunspot groups (25 July 1946, 5 April 1947, 17 May 1951 seen in CaII K1v). Courtesy Paris observatory.*

Extremely long filaments, such as those of figure 24 and figure 25, delineate giant magnetic cells on the Sun. Indeed, filaments are located above the inversion line of photospheric magnetic fields; large filaments can be considered as tracers of magnetic cells. Filaments of more than one solar radius length are not frequent and can last more than 6 rotations (figure 25). Here again, long term collections of observations, are essential to catch rare phenomena which contribute to a better understanding of solar physics.

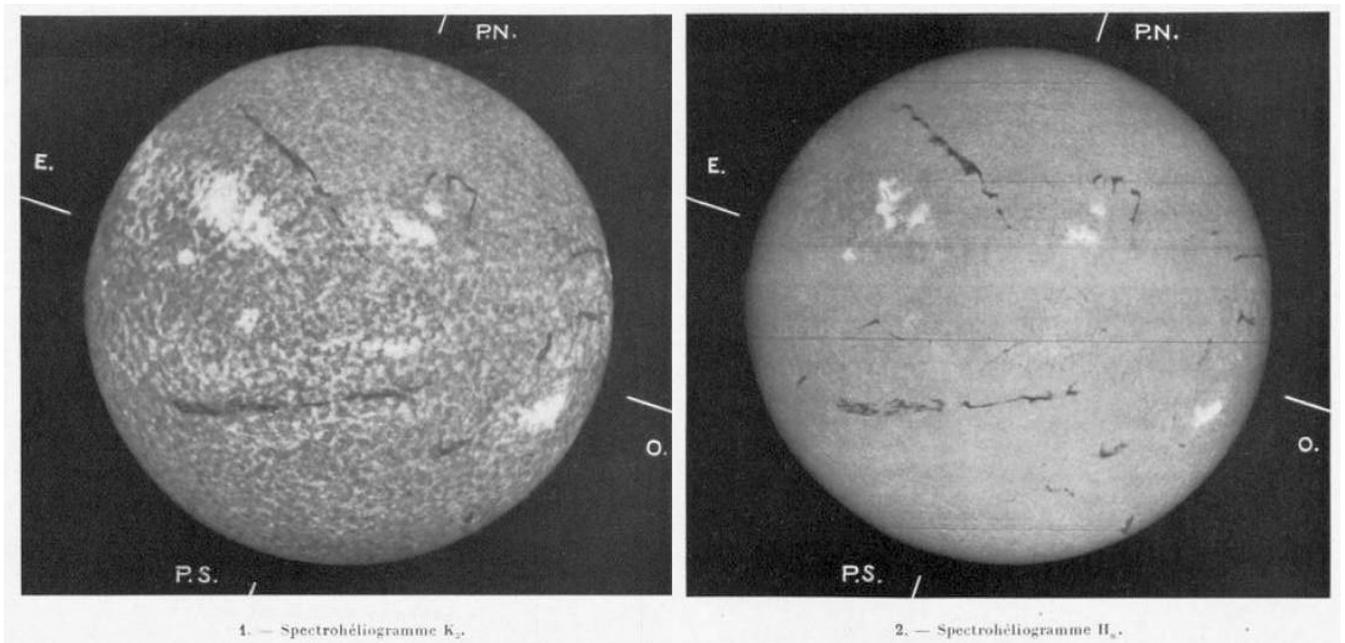

**Figure 24**. *Long filaments of 26 May 1930 in CaII K3 (left) and Hα (right). After d'Azambuja & d'Azambuja (1948).*

Fast solar activity events, such as flares, "Disparitions Brusques" of filaments, prominence eruptions and mass ejections are difficult to observe with a spectroheliograph, because of the moderate temporal resolution. For that reason, Hα filters were used after 1950. At Meudon, high cadence observations (1 minute) of dynamic events started in 1954 with Lyot filters, in particular in the frame of the International Geophysical Year (1957). However, until this date, spectroheliographs were the only available instruments to study fast evolving phenomena, and a few exceptional happenings were recorded with a maximum of twelve spectroheliograms for an event lasting a few hours. It was indeed a complicated task at the epoch of glass photographic plates. The prominence eruption of 18 June 1925 (figure 26 and figure 27) is one of the best events available in the collection. A "section" spectroheliogram (figure 27) was even produced to estimate the radial velocities (i.e. the velocity component orthogonal to the sky plane). Of course, prominence eruptions are today commonly observed in HeII 304 Å (the Lyα line of HeII at 80000 K) with the AIA telescope onboard SDO/NASA since 2010 (45 s cadence), but it is also the case with the EIT telescope onboard SOHO/ESA/NASA since 1996 (with a lower cadence and spatial resolution than SDO).

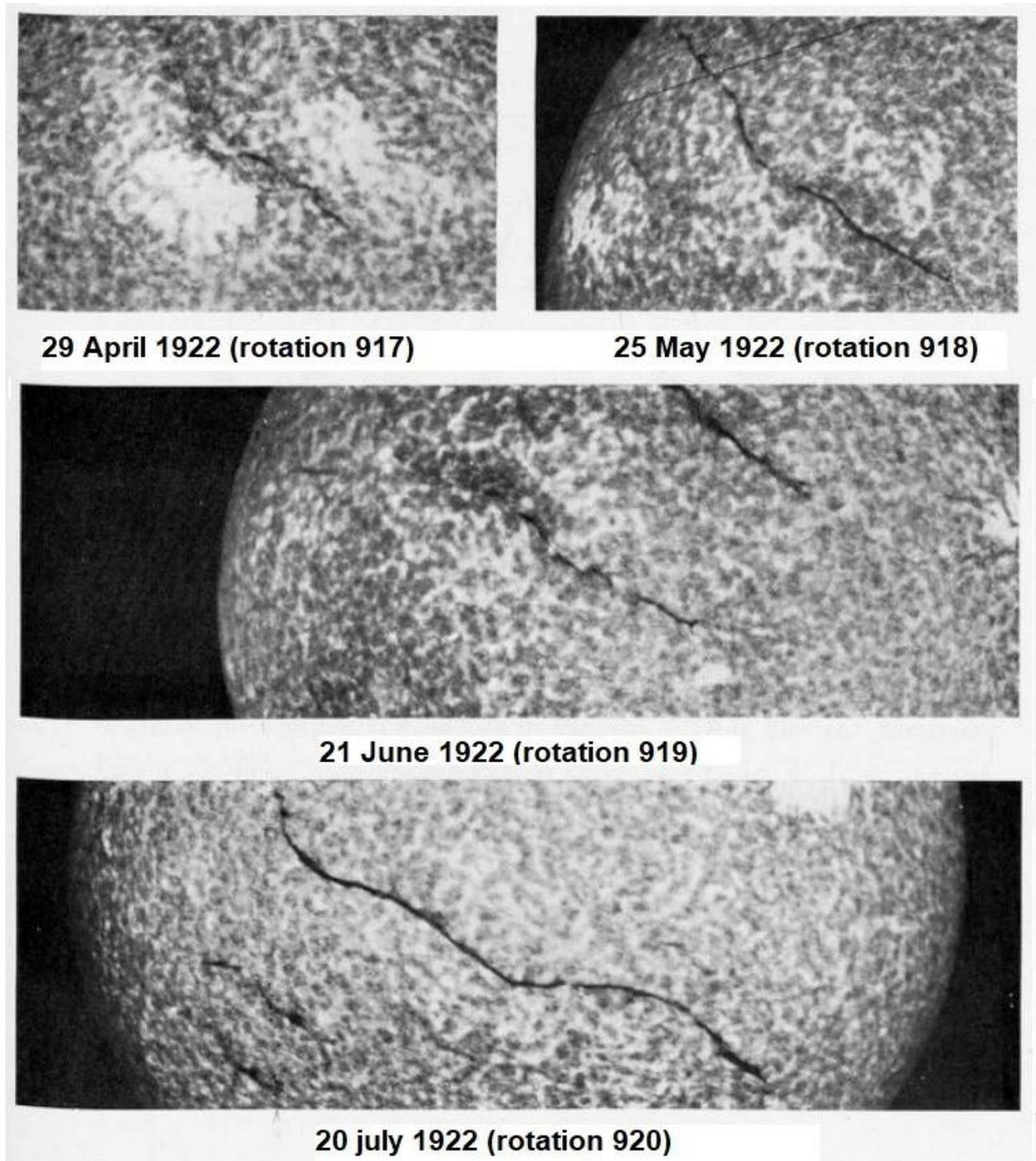

**Figure 25**. *An extremely long filament observed during several rotations. It started to form during rotation 915 (not shown). The length at rotation 920 exceeds one million kilometres. After d'Azambuja & d'Azambuja (1948).*

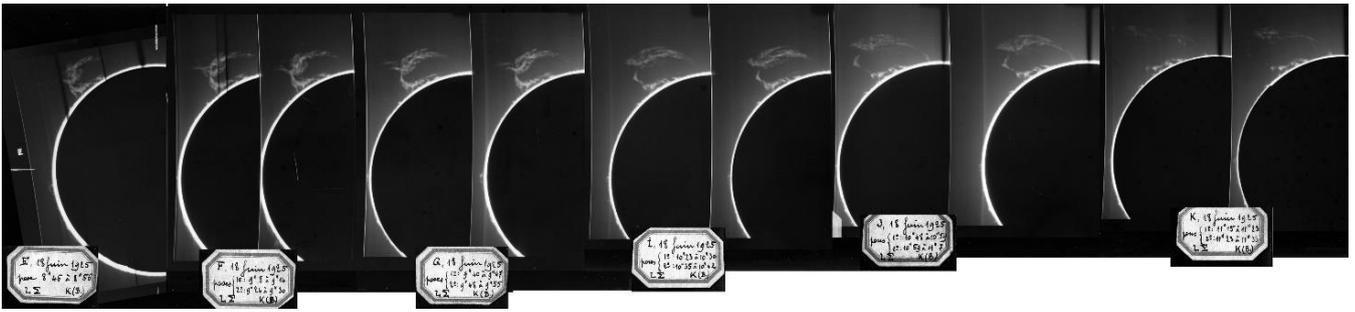

**Figure 26**. *Instability of a solar prominence at the limb observed in CaII K3 on 18 June 1925 (08:46 to 11:23) with Meudon spectroheliograph; an artificial moon was masking the solar disk. Courtesy Paris observatory.*

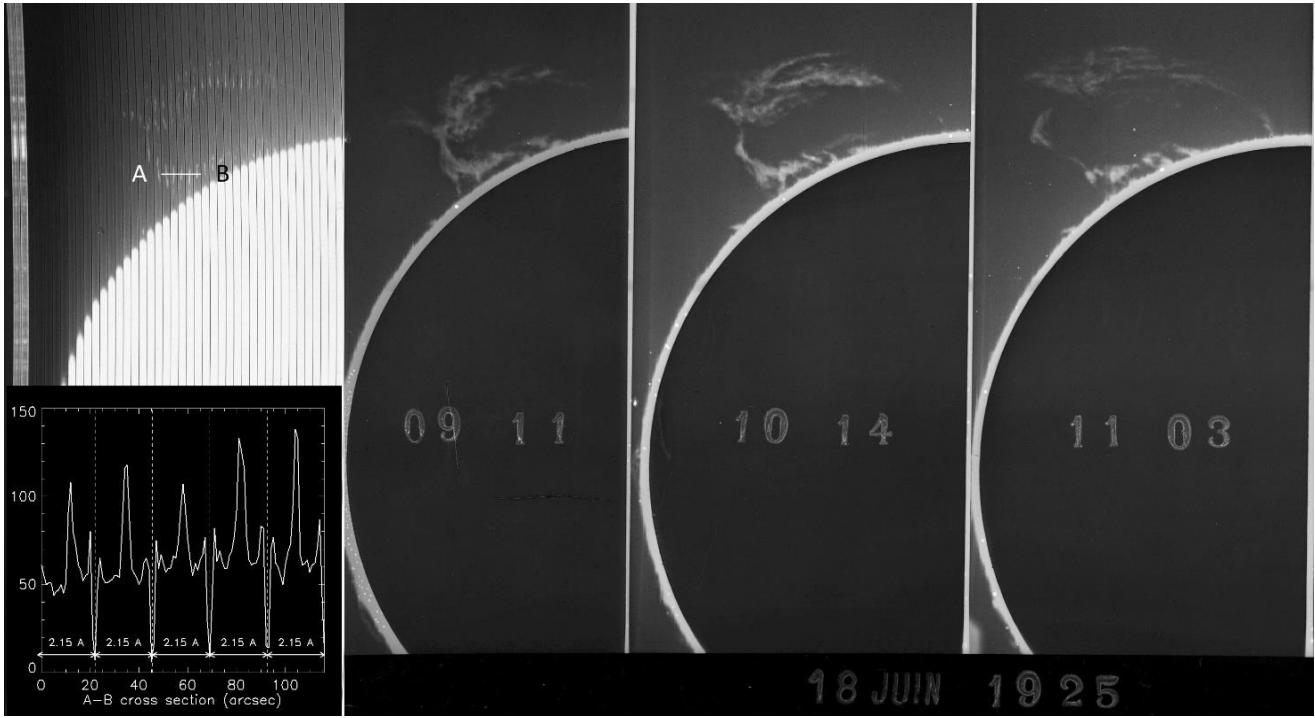

**Figure 27**. *Instability of the solar prominence of 18 June 1925 in the CaII K line. The image at left is a "section" spectroheliogram, obtained by moving the entrance slit by steps of 22". The five line profiles of CaII K (2.15 Å wide) at the bottom are those of the cross section AB. Courtesy Paris observatory.*

Figure 28 displays one of the "treasures" of the 10 solar cycles Meudon collection, this is the eruption of a huge solar prominence observed just after WW1 in May 1919 (observations were interrupted during the war). Observations were performed in the CaII K3 line with an artificial moon superimposed upon the solar disk to allow long exposure time.

The July 1946 group of figure 23 produced the strong solar flare of figure 29, and a huge geomagnetic storm. This phenomenon is one of the most energetic eruptions ever observed. André Danjon, the director of Paris-Meudon observatories, was there during this event and was fond of solar observations. However, Gualtiero Olivieri, who was observer, reported that "*his advices and commands given to the observers did not helped them to stay quiet and keep cool*" ! Flares originate in active regions and have a magnetic origin. The magnetic energy is stored in sunspot groups, such as the big one of figure 23; it is proportional to the surface and to the square of the magnetic field intensity. Unstable configurations often occur near the solar maximum and just after, and trigger huge flares. The 1946 event illustrates the importance of historical data. It allowed to predict the maximum energy of a solar flare, about $5 \cdot 10^{26}$ Joule (10 times the one ever measured by satellites of the modern era). The average energy of solar eruptions lies around $10^{25}$ Joule.

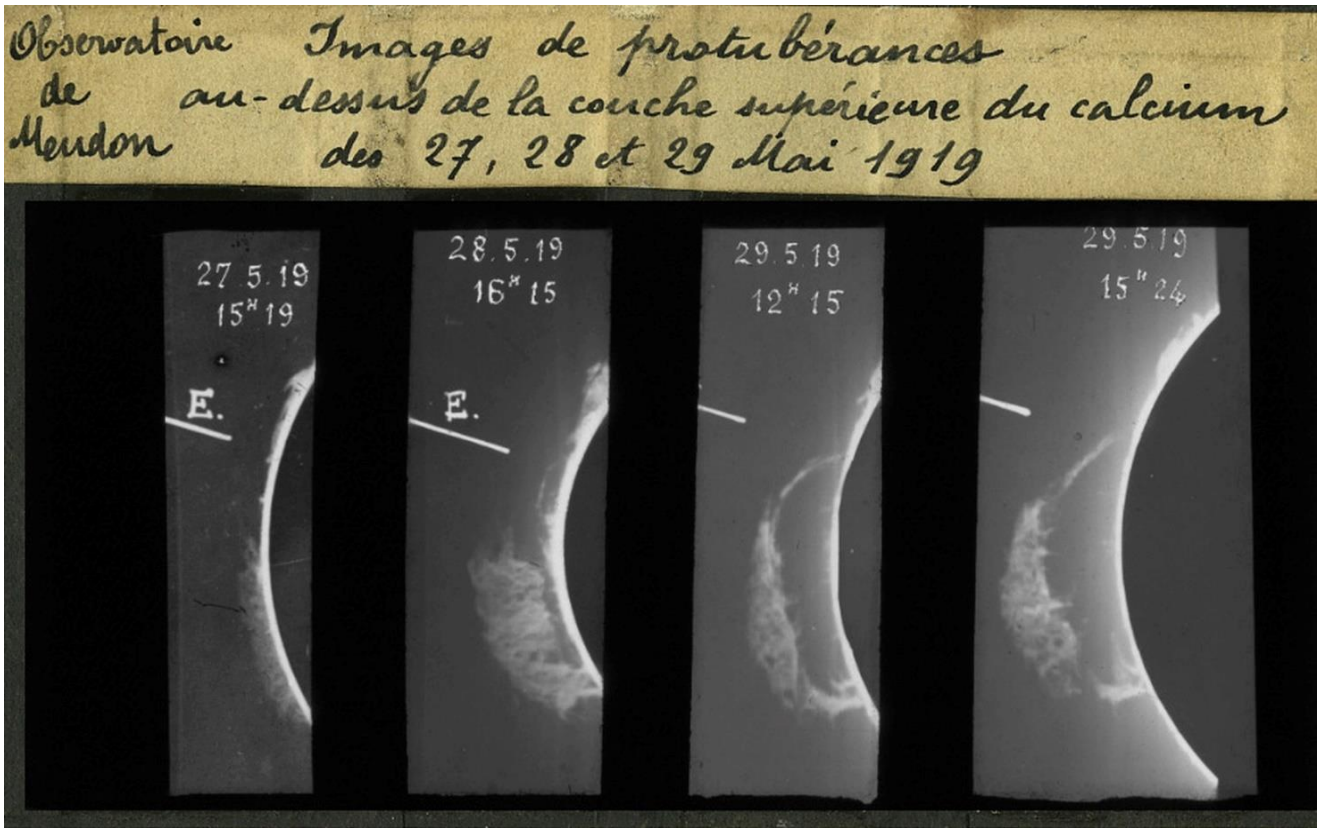

**Figure 28**. *Instability of the solar prominences in CaII K3 on 27, 28 and 29 May 1919 observed with Meudon spectroheliograph. The times are indicated in the picture. Courtesy Paris observatory.*

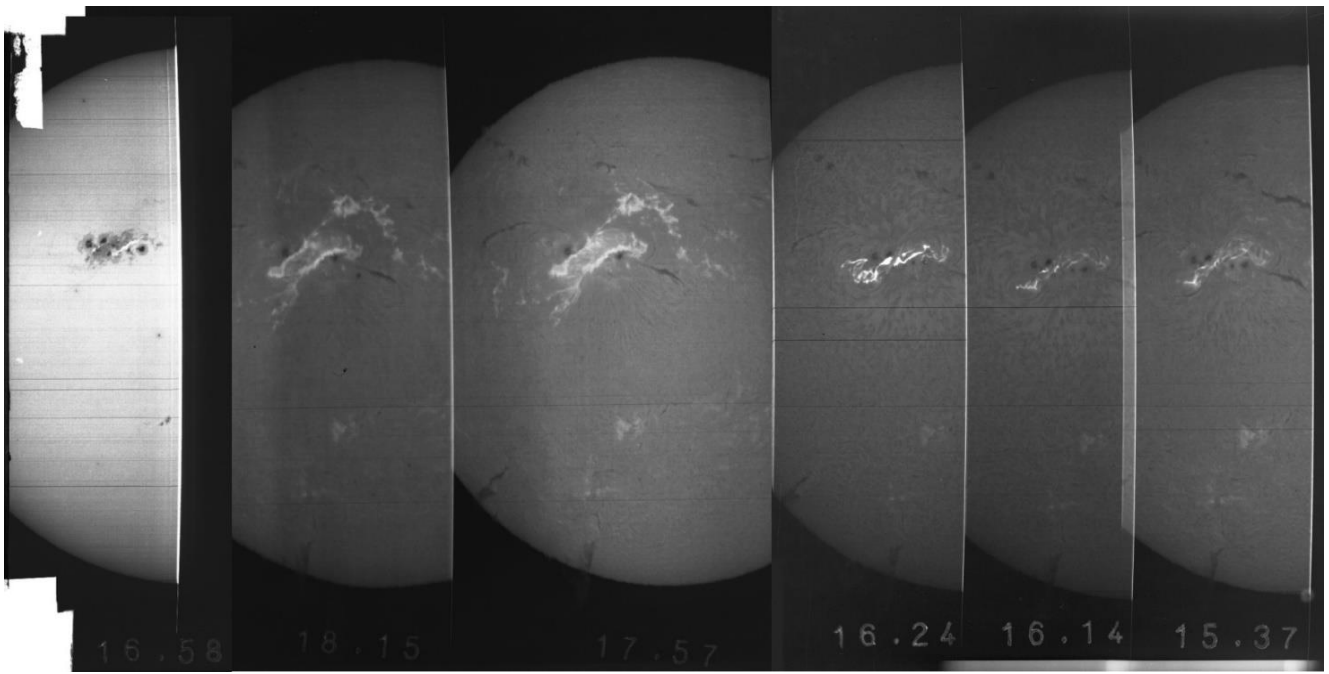

**Figure 29**. *The major two-ribbon solar flare of 25 July 1946 observed in Hα with Meudon spectroheliograph (the times are, from the right to the left, 15:37, 16:14, 16:24, 17:57, 18:15 UT). The left image at 16:58 was got in the HeI D3 5876 Å line. The short-duration flash phase (lasting minutes) occured at 16:14. Two bright ribbons (17:57, 18:15) formed during the long-duration gradual phase (lasting hours) and corresponded to the impact of energetic particles on the chromosphere. Courtesy Paris observatory.*

## 5 – CONCLUSION

Meudon spectroheliographs produce monochromatic images of the chromosphere in CaII K3 and Hα, and of the photosphere in the blue wing (K1v) of the CaII K line, on a daily basis since 1908. The collection extends over 10 solar cycles and contains more than 100000 spectroheliograms. However, between 1919 and 1954, a lot of special and unique observations were produced by Mr and Mrs d'Azambuja with the large 7-metres spectroheliograph dedicated to research. Many unusual spectral lines were studied, such as photospheric lines of FeI, CaI, HeI or SrII, and chromospheric lines of infrared CaII or HeI. Some observations, such as HeI 10830 Å, were the first world-wide, and showed that the initial choice of CaII K3, K1v and Hα was the best for imaging spectroscopy of the photosphere and the chromosphere. Besides, many dynamic and rare events were recorded with the spectroheliographs dedicated to systematic observations, such as huge flares or prominence ejections. Of course, spectroheliographs were not designed for high cadence phenomena, as their goal was low cadence and long term observations of the solar cycles. For that reason, instruments using Lyot Hα filters were used at Meudon after 1954 to record fast evolving events. During this survey, about 6 million images were registered on 35 mm films (130 kilometers !) until 1997.

## 6 - ON-LINE MATERIAL (VIDEO CLIPS)

Two examples of Hα and CaII K line scans with the 2017 numerical version of the Meudon spectroheliograph:

Halpha.mp4: spectroheliograms at various wavelength positions along he Hα line profile. The line core forms in the chromosphere (1500 km) and reveals dark filaments and bright plages. The line wings form below, in the photosphere, and shows sunspots. Spectral resolution 0.155 Å.

CaK.mp4: spectroheliograms at various wavelength positions along he CaII K line profile. The line core (K3) forms in the chromosphere (1900 km) and reveals dark filaments and bright plages. The line wings (K1v) form below, in the photosphere (500 km), and shows sunspots and facular areas. Spectral resolution 0.093 Å.

## 7 - ACKNOWLEDGEMENTS

The author thank I. Bualé and F. Cornu for spectroheliograms and archives of the Meudon solar collection.

## 8 - NOTES

**[1] The solar atmosphere**. It is composed of three layers: (1) the photosphere (the visible surface, temperature decreasing from 6000 K to 4500 K in 300 km) with dark sunspots and bright faculae around, (2) the chromosphere above (temperature increasing from 4500 K to 8000 K in 2000 km) with dark filaments and bright plages (corresponding to faculae in the photosphere), (3) the ionized and hot corona (2 million K). The corona extends at million kilometres and gives rise to the solar wind (charged particles), which propagates into the interplanetary medium. The chromosphere and the corona require spectroscopic means to reveal their structures, respectively via absorption lines of the visible spectrum or emission lines of the ultraviolet spectrum. The solar atmosphere follows a 11-years activity cycle and a 22-years magnetic cycle. Flares and coronal mass ejections occur in active regions a few years around the solar maximum; the maximum of the current cycle (number 25) is forecasted for 2025. The symbol K above (Kelvin) is the unit of the absolute temperature, related to the Celcius temperature by $T(K) = T(°C) + 273.15$.

**[2] The wavelength**. Electromagnetic waves, such as the visible light, are detected by the telescopes. They form a continuous spectrum (such as the rainbow) with superimposed spectral lines, revealing the presence of atoms, ions or molecules. In the visible or ultraviolet spectrum, the lines are identified by their wavelength, measured in nanometres (1 nm = $10^{-9}$ m), but spectroscopists often prefer the Angström (1 Å = 0.1 nm = $10^{-10}$ m). The visible solar spectrum (4000 Å - 7000 Å) reveals thousands of absorption lines, while emission lines occur in the ultraviolet spectrum (below 4000 Å) or at the limb.

 **[3] The solar structures**. Dark sunspots are regions of intense magnetic fields (0.1 - 0.3 T). Bright faculae or plages form, together with sunspots, active regions, and exhibit smaller magnetic fields (0.01 - 0.05 T). Dark filaments, also called prominences when seen at the limb, are thin and high structures (50000 km) of dense material suspended in the corona by weak magnetic fields (0.001 T). The symbol T is the unit of magnetic field (Tesla); the Gauss (1 G = $10^{-4}$ T) is also used. Flares occur in zones of unstable fields; reconnections convert magnetic energy into kinetic energy (ejections), radiation (X-rays) and heat (brightenings).

**[4] The Doppler effect**. When a light source moves in the observer's direction, the spectral lines are shifted towards shorter wavelengths (blueshift). On the contrary, if the source moves in the opposite direction, lines are shifted towards longer wavelengths (redshift). The Dopplershift w(Å) is a wavelength shift, proportional to the projection V of the velocity vector along the line-of-sight (it is also called radial velocity, positive towards the observer in solar physics): w = - λ (V / C), where λ is the line wavelength (Å) and C is the speed of light (3 $10^5$ km/s).

## 9 - REFERENCES

## 10 - THE AUTHOR

Dr Jean-Marie Malherbe, born in 1956, is astronomer at Paris-Meudon observatory. He got the degrees of "*Docteur en astrophysique*" in 1983 and "*Docteur ès Sciences*" in 1987. He first worked on solar filaments and prominences using multi-wavelength observations. He used the spectrographs of the Meudon Solar Tower, the Pic du Midi Turret Dome, the German Vacuum Tower Telescope, THEMIS (Tenerife) and developed polarimeters. He proposed models and MHD 2D numerical simulations for prominence formation, including radiative cooling and magnetic reconnection. More recently, he worked on the quiet Sun, using HINODE (JAXA), IRIS (NASA) and MHD 3D simulation results. He is responsible of the Meudon spectroheliograph since 1996.